\documentstyle[twoside,12pt]{article}
\input{amssym.def}
\input{amssym}

\newtheorem{thm}{Theorem}[section]
\newtheorem{defn}[thm]{Definition}
\newtheorem{lemma}[thm]{Lemma}
\newtheorem{prop}[thm]{Proposition}
\newtheorem{cor}[thm]{Corollary}

\newcommand{\TLm} {T_{L}^{m}}
\newcommand{\TLn} {T_{L}^{n}}
\newcommand{\TLms} {(T_{L}^{m})^{*}}
\newcommand{\LLm} {\Lambda_{L}^{m}}
\newcommand{\LLn} {\Lambda_{L}^{n}}
\newcommand{\LLms} {(\Lambda_{L}^{m})^{*}}
\newcommand{\Linf} {\Lambda_{\infty}}
\newcommand{\PLm} {\Pi_{L}^{m}}
\newcommand{\NLmn} {N_{L,m}^{(n)}}
\newcommand{\mLm} {\mu_{L}^{m}}
\newcommand{\mbm} {\breve{\mu}^{m}}
\newcommand{\mbone} {\breve{\mu}}
\newcommand{\Ltwok} {L_{(k)}^{2} (\TLm)}
\newcommand{\fLmt} {{\phi}_{L}^{m}[t]}
\newcommand{\finft} {{\phi}_{\infty}[t]}
\newcommand{\Tex} {T(\eta,\xi)}
\newcommand{\Tkex} {T^{(k)}(\eta,\xi)}
\newcommand{\R} {{\Bbb R}}
\newcommand{\Z} {{\Bbb Z}}
\newcommand{\N} {{\Bbb N}}
\newcommand{\Id} {{\bf 1}}
\newcommand{\pha} {\phantom{=}}
\newcommand{\qed} {\hfill {\sc Q.E.D.}}
\newcommand{\fn} {function}
\newcommand{\erg} {ergodic}
\newcommand{\mix} {mixing}
\newcommand{\Fou} {Fourier}
\newcommand{\cs} {coherent state}
\newcommand{\td} {thermodynamical limit}
\newcommand{\psd} {pseudo-differential}
\newcommand{\tdlim} {\lim_{m,L\to\infty \atop m/L \to\rho}}
\newcommand{\la} {\langle}
\newcommand{\ra} {\rangle}
\newcommand{\ei}[1] {e^{2\pi i {#1}}}
\newcommand{\emi}[1] {e^{-2\pi i {#1}}}
\newcommand{\Op} {{\rm Op}}
\newcommand{\Tr} {{\rm Tr}_{L,m,k}\,}
\newcommand{\Hm} {{\cal H}_{m}}
\newcommand{\OL} {{\cal O}(L^{-\infty})}
\newcommand{\Sm} {{\Bbb S}^{m}}
\newcommand{\sectcount} {\setcounter{equation}{0}}

\renewcommand{\b} {\!\!}
\renewcommand{\a} {\big|}
\renewcommand{\c} {\,\sharp\,}
\renewcommand{\hat} {\widehat} 
\renewcommand{\tilde} {\widetilde} 

\begin{document}

\title{
	Ergodic Properties of the Quantum Ideal Gas in the 
	Maxwell-Boltzmann Statistics
}
\author{
	Marco Lenci \thanks{On leave at: Mathematics Department, Princeton 
	University, Princeton, NJ 08544, U.S.A. E-mail: 
	marco@math.princeton.edu} \\
	Dipartimento di Matematica \\
	Universit\`a di Bologna \\
	40127 Bologna \\
	Italy \\
}
\date{April 1996}

\maketitle

\begin{abstract}
	It is proved that the quantization of the Volkovyski-Sinai model 
	of ideal gas (in the Maxwell-Boltzmann statistics) enjoys at the 
	\td\ the property of quantum \mix\ in the following sense:
	\begin{displaymath}
		\lim_{|t|\to\infty} \tdlim \omega_{\beta,L}^{m}
		(e^{iH_{m}t/\hbar} A e^{-iH_{m}t/\hbar} \, B) =
		\tdlim \omega_{\beta,L}^{m}(A) \cdot
		\tdlim \omega_{\beta,L}^{m}(B).
	\end{displaymath}
	Here $H_m$ is the Schr\"odinger operator of $m$ free
	particles moving on a circle of length $L$; $A$ and $B$ are 
	the Weyl quantization of two 
	classical observables $a$ and $b$; 
	$\omega_{\beta,L}^{m}(A)$ is the corresponding quantum Gibbs state.
	Moreover one has
	\begin{displaymath}
		\tdlim \omega_{\beta,m} (A) = P_{\rho,\beta}(a)
	\end{displaymath}
	where $P_{\rho,\beta}(a)$ is the classical Gibbs measure.
	\par
	The consequent notion of quantum \erg ity is also independently 
	proven.
\end{abstract}
	
\section{Introduction.} \label{sec-intro}
\sectcount

The purpose of the present paper is to analyze the quantum \erg\ 
properties of the Volkovyski-Sinai model of ideal gas, quantized 
according to a Maxwell-Boltzmann statistics (i.e. all particles are 
distinguishable). This paper represents the companion paper of 
\cite{gm} where the same result is proved for a strongly analogous 
system, namely the infinite harmonic chain with suitable restrictions 
on the normal mode frequencies.
\par
These two systems provide examples of {\em kinematic quantum chaos}.
We borrow the expression {\em kinematic chaos} from the enlightening 
paper by \cite{jlp}. By that we mean trivial motion whose chaotic 
behavior is due to the randomness of the infinite-dimensional initial 
condition (see, besides \cite{jlp} itself, the remark after 
Theorem \ref{thm-erg}). For both the ideal gas and the 
harmonic chain, this is a classical feature of the system, which 
works on the quantum level as well because of the {\em exact Egorov 
Theorem} (Lemma \ref{lemma-q-evol-c}). The latter is a fundamental 
fact here, since it allows us to treat the quantum time evolution as the 
classical one.
\par
A clear explanation of the motivations for the investigation of the 
quantum behavior of infinite systems is given in \cite{gm}, as well 
as further references. Here we just sketch what the immediate problems 
are, concerning the search for {\em quantum chaos}.
\par
To fix the ideas, let $H$ be a self-adjoint operator acting on 
$L^{2}(\Omega), \Omega\in \R^{m}$, resulting from the quantization of 
some classical Hamiltonian \fn\ ${\cal H}$ over $\R^{m}\times\Omega$.  
Suppose, as it happens in all interesting cases, that $\sigma(H)$ is 
discrete.  Consider two operators $A,B \in {\cal L}(L^{2}(\Omega))$, 
which is regarded as our set of observables, and define $\Theta[t](A) := 
e^{iHt/\hbar} A e^{-iHt/\hbar}$, the Heisenberg evolution.  All the 
physical experiments one can do on such a system imply a certain 
``measure'' on the observables is used. In quantum mechanics such 
measures are {\em states} over the algebra of the operators. These 
{\em quantum ensembles} (\cite{r}, \S1.3) are typically described as
\begin{equation}
	\omega_{\varrho}(A) = \frac{{\rm Tr} \, A\, \varrho}
	{{\rm Tr} \,\varrho},
\end{equation}
with $\varrho$ a suitable trace-class operator. A suitable definition of 
\mix\ would be, then (\cite{br}, \cite{b}, \cite{gm})
\begin{equation}
	\lim_{|t|\to\infty} \omega_{\varrho}(\Theta[t](A),\, B) = 
	\omega_{\varrho}(A) \: \omega_{\varrho}(B),\ \ \ 
	\forall A,B \in {\cal L}(L^{2}(\Omega)).
\end{equation}
It is easily realized, writing $\omega_{\varrho}$ by means of the 
matrix elements of the operators, w.r.t. the eigenbasis of $H$, that 
such a property can never be verified, {\em for each underlying 
classical dynamics}.  The same is true for any reasonable definition 
of \erg ity, as Von Neumann's formula (\cite{vn}) shows (see \cite{gm} 
for details).  This is a consequence of the quasi-periodicity of the 
classical evolution -- as long as we have finite degrees of freedom -- 
and it is called ``quantum suppression of classical chaos'' 
(\cite{jlpc}, \cite{jlp}).  Hence the idea of taking the number of 
degrees of freedom to infinity.
\par
The system we consider is the quantization of the ideal gas in the 
formulation given in \cite{s}: i.e. $m$ non-interacting particles moving 
freely on a circle of length $L$, when $m$ and $L$ are taken to infinity, 
subject to the finite density requirement $m/L \to\rho$.
\par 
For an outline of the ``analytic approach'' we will follow in studying
the quantum infinite system in question, the reader is 
definitely referred to the introduction of the companion paper 
\cite{gm}, due to the similarities of the two works. There the 
consequences of such a study are also properly emphasized.
In the next paragraph we just point out the structural differences, 
between the two models, which require non obvious modifications of 
the arguments of \cite{gm} valid for the harmonic chain.
Specifically
\begin{itemize}

\item[(i)] The other important classical mechanism which provides the
unpredictability (mixing) of the time evolution here, besides the 
mentioned kinematic effect, is the {\em symmetry} of the 
observables under particle exchange. This corresponds to the physical 
fact that one is not able to distinguish between the particles in a gas.
Actually, such a restriction on the observables also has the noticeable 
outcome to allow the interchange of the time average limit with the \td\ 
(see Theorem \ref{thm-erg}, Corollary \ref{cor-erg}, and relation 
(\ref{thm-erg-1})). The remark after Theorem \ref{thm-erg} will 
contain more comments. On the quantum side, the simmetry of the 
observables would entail for a Bose-Einsten or a Fermi-Dirac statistics. 
Nevertheless, we use Maxwell-Boltzmann for the sake of convenience: see 
remark 2 in \S\ref{subs-quant}.

\item[(ii)] The phase space of any $m$-particle subsystem is 
$\R^{m}\times(L\,S^{1})^{m}$, with $L\,S^{1}$ denoting henceforth the 
circle of radius $L$. In other words the phase space has a cylindrical 
structure. This has the effect of making necessary a Bloch 
decomposition of $L^2(\R^{m})$, and consequently a direct fiber 
decomposition of operators, if we want to consider functions on phase 
space as symbols of operators on a one-to-one correspondence under 
quantization. Equivalently, $\R^{m}\times(L\,S^{1})^{m}$ generates 
a cylindrical Heisenberg group (see \cite{dbg} and \S\ref{subs-quant}) 
and its only faithful unitary representation is given by a fibered 
$L^2$ space.

\item[(iii)] The coherent states we use are those
adapted to the cylindrical phase space, as presented in  
\S\ref{appendix-cs} (refer also to \cite{p}, \cite{dbg}).

\item[(iv)] The infinite-particle limit is a true thermodynamic limit,
because here not only we have $m\to\infty$, but also $L\to\infty$ under
the constraint of finite density $\displaystyle \frac{m}{L}\to\rho,
\rho>0$.

\item[(v)] On the other hand in this model the classical dynamics is just
free motion. This entails a simplification: the Weyl symbol of the quantum
$m$-particle Schr\"odinger operator $H_m$ is obviously the classical 
Hamiltonian $\Hm = (1/2) \sum_{i=1}^{m} p_{i}^{2}$, namely $H_{m} = 
\Op(\Hm)$. Then the Weyl symbol of the (unnormalized here) quantum Gibbs 
measure is
\begin{displaymath}
	e^{-\beta H_{m}} = \Op (e^{-\beta\Hm}),
\end{displaymath}
where, here and above, $\Op := \Op^{W}$ denotes the operation of Weyl 
quantization of a symbol, whose definition in the present context is 
recalled in \S\ref{subs-quant} below.
\end{itemize}

Now, as it will be explained in sharper detail in \S\ref{subs-Gibbs}, 
consider the normalized quantum Gibbs state at inverse temperature 
$\beta$ (for a system of $m$ free particles in $L\,S^{1}$), namely the 
functional 
\begin{equation}
	\omega_{\beta,L}^{m}(A) := \frac{\mbox{Tr}\, A \, e^{-\beta H_{m}}}
	{\mbox{Tr}\, e^{-\beta H_{m}}} ,
\end{equation}
defined, for instance, over the class of bounded operators $A$ which 
are ``$L$-periodic in the coordinate variable'' (refer to 
\S\ref{subs-quant}). The application Tr is practically the trace over 
$L^{2}(\R^{m})$ (see \S\ref{subs-Gibbs} for details). Then the main 
result of the present work, in analogy with \cite{gm}, can be stated as
\begin{equation}
	\lim_{|t|\to\infty} \tdlim \omega_{\beta,L}^{m} \left( 
	\Theta^{m}[t] (A)\, B \right) = \tdlim \omega_{\beta,L}^{m}(A) 
	\cdot \tdlim \omega_{\beta,L}^{m}(B).
	\label{intro-mix}
\end{equation}
(see also Theorem \ref{thm-mix} and (\ref{thm-mix-1})). Here $A$ and 
$B$ are actually dependent on $L$ and $m$ and represent the operators 
quantizing two classical observables, $a$ and $b$, over the Hilbert 
spaces for $m$ particles in a $L$-circle (square integrable Bloch \fn s). 
Also $\Theta^{m}[t] (A) := e^{iH_{m}t/\hbar} A e^{-iH_{m}t/\hbar}$. 
Moreover
\begin{equation} 
	\tdlim \omega_{\beta,m}(A) = P_{\rho,\beta}(a).
\end{equation}
Formula (\ref{intro-mix}) is the quantum \mix\ property and induces a 
consistent formulation of the quantum \erg ity: for instance
\begin{equation}
	\lim_{T\to\infty} \tdlim \omega_{\beta,L}^{m} ( (\Xi^{m}[T] 
	(A))^{2}) = \tdlim ( \omega_{\beta,L}^{m}(A) )^{2},
\end{equation}
if $\Xi^{m}[T] (A) = (1/2T) \int_{-T}^{T} \Theta^{m}[t] (A) \,dt$.
\smallskip\par
The present model of quantum ideal gas is structurally different 
from the free Bose gas or the free Fermi gas in the grandcanonical 
ensemble, discussed e.g. in \cite{br}, \S5.2. The arguments 
(\cite{br}, \cite{b}) yielding the mixing property with respect to the 
KMS states through the asymptotic abelianess of the CCR (or CAR) 
automorphisms generated by free dynamics do not apply in this context.
\par 
The paper is organized as follows: in the next section we briefly recall 
the ideal gas model and construct its quantization; in \S\ref{sec-statement} 
we state the results, whose proofs are given in \S\ref{sec-proofs}. 
Finally, in the appendix we recall the construction of the coherent 
states on the cylinder (obtained e.g. in \cite{dbg}) with some 
additional details; we collect two technical lemmas; and we exploit the 
properties of the convolution over $L\,S^{1}$, as $L\to\infty$, which 
are crucially used in the proofs.

\section{The Classical Ideal Gas and Its Quantization.} 
\label{sec-ideal-quant}
\sectcount

\subsection{The Volkovyski-Sinai model.} \label{subs-classical}

To make the exposition self-contained, a brief reminder is here given of 
the Volkovyski-Sinai model of ideal gas. The reader is referred to
\cite{s}, Lecture 8, for details.
\par
Consider a system of $m$ free particles of unit mass constrained to move 
on a circle of length $L$. Its Hamiltonian \fn\ is
\begin{equation}
	\Hm(p,q) = \frac{1}{2} \sum_{j=1}^{m} p_{j}^{2} = p^{2}/2
\end{equation}
defined on the phase space $\LLm := \R^{m} \times (L\,S^{1})^{m}$ which,
with a view to the limiting case $L\to\infty$ to be considered below, will
be identified with $\R^{m} \times \TLm,\ \TLm := [-L/2,L/2)^{m}$. The
physical intuition is to stretch the circle $L\,S^{1}$ more and more
towards a straight line, namely $\R$. 
\par
On $\LLm$ the motion is given by the flow map: $\fLmt(p,q) := (p,q+pt)$.
The dynamical system is completely integrable and $\LLm$ is decomposed in
a non-countable family of invariant tori: all motions are quasi-periodic.
Introducing any ``reasonable" measure on $\LLm$ (in \cite{s} the
microcanonical measure is used; for reasons which will become clear when
quantizing, the natural measure to introduce here is the canonical one)
the system is of course not even \erg. 
\par
But a suitable \td\ of it might be. The construction of the infinite 
dynamical system is based on the idea that the particles should be 
``unlabeled'' the particles. More precisely, the phase space for the 
infinite system is defined as follows:
\begin{equation}
	\Linf := \{ (p,q);\ q \mbox{ countable subset of }\R,\: p:q \to\R \},
	\label{Linf}
\end{equation}
with the interpretation that $q = \{x_{1},x_{2},\ldots,x_{j}, \ldots \}$ 
contains the positions of the particles, now undistinguishable, and 
$p$ is a \fn\ such that $p(x_{j})$ gives the velocity of the particle 
located at $x_{j}$. It may occur that more than one particle -- say 
$n$ -- are located at $x_{j}$: in this case, with an abuse of notation with 
respect to definition (\ref{Linf}), $p(x_{j})$ is an $n$-tuple of 
velocities. The flow $\finft: \Linf\to\Linf$ is defined accordingly: 
$\finft(p,q) = (p',q')$, where $q' := \{ x+p(x)t;\: x\in q\}$ and 
$p':q'\to\R$ is $p'(x+p(x)t) := p(x)$. 
\par
What we have introduced is the natural limiting object of the spaces 
$\LLm/\Sm$, $\Sm$ being the group of permutations of $m$ coordinates 
and momenta. Such spaces are expressly 
defined in \cite{s} in a way completely analogous to (\ref{Linf}), 
i.e. as the collection of all couples $(p,q)$ with $q\subset 
L\,S^{1}, \, \# q\leq m$ and $p:q \to\R$ having $m$ values, in the sense 
specified above. They can be regarded as belonging to $\Linf$, since
$q \subset L\,S^{1} \simeq [-L/2,L/2)$. This motivates the above 
identification.
\par
This remark enables us to consider \fn s defined  
on $\Linf$ as having a natural restriction on $\LLm/\Sm$. And a \fn\ 
$f$ on $\LLm/\Sm$ is simply a totally symmetric \fn\ on $\LLm$, 
namely $f \circ \PLm$, where $\PLm$ is the natural immersion $\LLm 
\to \Linf$:
\begin{equation}
	\PLm(p_{1},\ldots,p_{m},q_{1},\ldots,q_{m}) := (p:q\to\R,q:=
	\{ q_{1},\ldots,q_{m} \} ),
\end{equation}
$p$ taking values $p(q_{1}) = p_{1}, \ldots, p(q_{m}) = p_{m}$.
It will be useful in the remainder to notice that
\begin{equation}
	\finft \circ \PLm = \PLm \circ \fLmt
	\label{comm-of-flows}
\end{equation}
To complete the definition of our infinite dynamical system, we need 
to specify the measurable \fn s, i.e. to fix a $\sigma$-algebra 
${\cal A}$ and -- after that -- a probability measure on it. The 
answer in \cite{vs} is ${\cal A} := \sigma(\gamma(\Delta) 
)_{\Delta\in{\cal B}(\R)}$.  Here $\Delta$ runs among the Borel sets 
in $\R$ and $\gamma(\Delta)$ is the $\sigma$-algebra of all subsets of 
$\Linf$ depending only on the positions and momenta of the {\em 
unlabeled} particles in $\Delta$.
\footnote{More precisely, $\gamma(\Delta)$ is the
$\sigma$-algebra for which the measurable \fn s are all functions $f$ 
on $\Linf$ depending only on values taken by the particles in $\Delta$
{\em and} measurable when viewed as \fn s on $\LLm$:
$f \circ \PLm$.}
For the sake of simplicity, we just restrict ourselves to real \fn s.
\par
Examples of measurable \fn s are: $f_{\Delta}(p,q) := \# (q\cap\Delta)$, 
the number of particles of the configuration $q$ located in $\Delta$; or 
$g_{\Delta}(p,q) := \sum_{x\in\Delta} p(x)$, the total momentum of the 
particles in $\Delta$.
\par
We endow ${\cal A}$ with the measure $P_{\rho,\beta}$ defined by the 
following properties: 
\begin{enumerate}
\item 
The distribution of particles in the configuration space is 
Poissonian with parameter $\rho$; that is
\begin{equation}
	P_{\rho,\beta} (\{(p,q);\: f_{\Delta}(p,q) = n\}) = e^{-\rho|\Delta|}
	\frac{(\rho|\Delta|)^{n}}{n!};
	\label{Poisson}
\end{equation}
for every $\Delta\in{\cal B}(\R)$. This implies that the distributions 
over two disjoint Borel sets $\Delta_{1}$ and $\Delta_{2}$ are independent.
\item 
The momenta are independent centered Gaussian variables with variance 
$1/\beta$. This means that, fixed $A\in {\cal B}(\R^{n})$ and a vector 
$(x_{1},\ldots,x_{n}) \in\R^{n}$, we have
\begin{equation}
	P_{\rho,\beta} ( \{ (p(x_{1}),\ldots,p(x_{n})) \in A \} \:\a\: 
	\{ \, \{x_{1},\ldots,x_{n}\}\subset q \} ) = \int_{A} \left[ 
	e^{-\beta p^{2}/2}dp \right]_{N}
	\label{Maxwell-distr}
\end{equation}
$[\cdots]_{N}$ being the normalized measure. This is a Maxwell 
distribution with inverse temperature $\beta$. 
\end{enumerate}
This measure has been chosen intentionally as the limit of the 
canonical measures over the phase spaces of finite numbers of 
particles. 
\begin{prop}
	Let $a$ measurable in ${\cal A}$. At the \td, i.e. $m,L \to\infty; 
	m/L \to\rho$,
	\begin{displaymath}
		\int_{\LLm} (a\circ\PLm)(p,q) \left[ e^{-\beta p^{2}/2}dpdq 
		\right]_{N} \longrightarrow P_{\rho,\beta}(a).
	\end{displaymath}
	\label{limit-of-meas}
\end{prop}
As regards the proof, the statement concerning the distribution 
of the positions is clearly explained in \cite{s}, while 
(\ref{Maxwell-distr}) trivially holds since the distributions over the
momentum spaces for a finite number of particles are given as Maxwellian 
with inverse temperature $\beta$. 
\par
We conclude this section formulating the key
\begin{thm}
	{\rm\cite{vs}} The measure $P_{\rho,\beta}$ is invariant under $\finft$ 
	and the dynamical system $(\Linf,{\cal A},P_{\rho,\beta},\finft)$ is 
	a K-flow.
	\label{vs-thm}
\end{thm}
\smallskip 
It may be worth noticing here that the infinite dynamical system just 
recalled has a nice abstract construction, which is described in 
\cite{gla} (and shortly in \cite{b}, Example 2.34). It is called the 
{\em Poisson system} constructed for a one-dimensional free particle 
with a Maxwell velocity distribution. Its ergodic properties follow 
via a general technique, namely the {\em Bernoulli construction}.  
This view-point shows clearly that ${\cal A}$ is generated by the 
following sets:
\begin{equation}
	B_{\Delta,\Gamma}^{(n)} := \big\{ (p,q)\in\Linf \:\a\: \# \{ x\in q 
	\,|\, x\in\Delta, p(x) \in\Gamma \} = n \big\},
	\label{B-Delta,Gamma}
\end{equation}
where $\Delta,\Gamma \in {\cal B}(\R)$. Such a remark will we useful 
while proving the statements. 
\par
Nonetheless the approach we have chosen has the advantage of 
constructing the infinite-particle dynamical system through a \td\ 
(see above, and in particular Proposition \ref{limit-of-meas}).  This 
fact will be crucial throughout this paper.

\subsection{The quantization} \label{subs-quant}

The Hilbert spaces associated to a quantum system of $m$ particles on 
a circle of length $L$ are denoted by $\Ltwok, k\in [0,1/L)^{m}$. Each 
of these is defined as
\begin{equation}
	\Ltwok := \big\{ f \mbox{ on }\R^{m} \a \forall j\in\Z^{m}, f(q+Lj) = 
	\ei{Lk\cdot j} f(q), \int_{\TLm}\b |f|^{2}\b <\infty \big\},
	\label{def-Ltwok}
\end{equation}
that is the space of the Bloch \fn s of parameter $k$ which are 
square-integrable on a given fundamental domain. Concerning this 
definition, we remark:
\begin{enumerate}
\item
The above family of spaces is the familiar one for 
Schr\"odinger operators with periodic potential (see
\cite{rs}, \S XIII.16).  They have to be simultaneously considered for 
all values of $k\in [0,1/L)^{m}$, otherwise the quantization 
application is not well defined since the Schr\"odinger 
representation of the Heisenberg group is not faithful: e.g., in one 
dimension, if we selected only $k=0$, then $T(L,0) = T(0,0) = \Id$ 
(see below).
\par
As a matter of fact, the whole Hilbert space $L^{2}(\R^{m})$ is 
recovered through the standard direct integral formula (\cite{rs})
\begin{equation}
	\int_{[0,1/L)^{m}}^{\oplus} \Ltwok dk \simeq L^{2}(\R^{m}).
	\label{Bloch-dec}
\end{equation}
More details are given in \S\ref{appendix-cs} and in \cite{dbg}.
A basis for $\Ltwok$ is obviously $e_{\alpha}^{(k)}:= L^{-m/2}
\ei{(\alpha+k)\cdot x}$, with $\alpha\in (\Z/L)^{m}$.
\item 
The choice of the Hilbert spaces in (\ref{def-Ltwok}) 
corresponds to a Maxwell-Boltzmann statistics, since we ask for no 
wave \fn\ symmetry with respect to particle permutations.  One could 
also think of quantizing this system according to a Bose-Einstein or a 
Fermi-Dirac statistics.  To give a physical explanation, we are 
considering particles which are in principle enumerable but our 
observables do not see this enumeration.  This approximation makes 
sense in the semiclassical realm.
\footnote{On the other hand, we stress that this paper does not 
involve any kind of semiclassical limit.}
\end{enumerate}
A further step towards the definition of the quantization 
application is the introduction of the \Fou\ transform and 
antitransform in $\LLm$. We define first the dual of the phase space: 
$\LLms := \R^{m}\times\TLms$ where $\TLms := (\Z/L)^{m}$. Now, if 
$b\in {\cal S}(\LLm)$, that is the Schwartz class of \fn s in 
$\LLm$, then for $(\eta,\xi)\in\LLms$, the \Fou\ transform of 
$b$ is
\begin{equation}
	\hat{b}(\eta,\xi) := \int_{\R^{m}} \int_{\TLm} b(p,q)\, 
	\emi{(\eta\cdot p + \xi\cdot q)} dq dp.
\end{equation}
Correspondingly, the antitransformation is given by
\begin{equation}
	b(p,q) := \frac{1}{L^{m}} \sum_{\xi\in\TLms} \int_{\R^{m}} 
	\hat{b}(\eta,\xi)\, \ei{(p\cdot\eta + q\cdot\xi)} d\eta.
	\label{Fou-antitransf}
\end{equation}
The Heisenberg group to be considered in this situation is the 
naturally induced {\em cylinder subgroup} of the Heisenberg group on 
$\R^{2m}\times\R$, namely $\LLms\times\R$ endowed with the product law
\begin{equation} 
	(\eta,\xi,\tau)(\eta',\xi',\tau') = (\eta+\eta,\xi+\xi',
	\tau+\tau' +\frac{1}{2}(\eta\cdot\xi' - \xi\cdot\eta')).
\end{equation}	
Accordingly (see \cite{fo}), its unitary Schr\"odinger representation 
in $L^{2}(\R^{m})$ is defined in the following way:
\begin{equation}
	(T(\eta,\xi)f)(x) = \ei{\xi\cdot(\eta/2 + x)} f(x+\eta)
\end{equation} 
This can be formally written as $T(\eta,\xi) = \ei{(\eta\cdot P + 
\xi\cdot Q)}$, where $Q$ corresponds to the multiplication operator by 
$q$, and $P = (2\pi i)^{-1} \nabla_{x}$.  We have therefore taken 
$\hbar = (2\pi)^{-1}$.  It is evident that $T(\eta,\xi)$ preserves 
$\Ltwok$: we denote $T^{(k)}(\eta,\xi)$ its restriction to that space.
\par
We are now in position to define the quantization application.
If $b$ is a \psd\ symbol of a given finite order (again \cite{fo} or 
\cite{sh}) on $\LLm$, then
\begin{equation}
	\Op(b) := \frac{1}{L^{m}} \sum_{\xi\in\TLms} \int_{\R^{m}} 
	\hat{b}(\eta,\xi)\, T(\eta,\xi) \,d\eta;
	\label{Opb1}
\end{equation}
with $\hat{b}$ possibly interpreted in distributional sense.  
\par
The restriction of this operator to the invariant space $\Ltwok$ will be 
once more denoted by $\Op(b)^{(k)}$. The above definition is nothing 
else than the standard Weyl quantization induced by the cylindrical 
Heisenberg group and subject to our choice of inverse Fourier transform 
(\ref{Fou-antitransf}). As a matter of fact, elementary algebraic  
manipulations yield the following explicit formula: for 
$f^{(k)}\in \Ltwok$,
\begin{equation}
	\left( \Op(b)^{(k)}f^{(k)} \right)(x) = \int_{\R^{m}} \b
	\int_{\R^{m}} b \left(\frac{x+y}{2},p\right) \ei{p \cdot (x-y)} 
	f^{(k)}(y) dy dp.
	\label{Opb2}
\end{equation}
Remark that, since $f^{(k)} \not\in {\cal S}(\R^{2m})$, a priori 
so this makes no sense even as an oscillatory integral. The sense we 
can give it is, again, distributional, as we know (\cite{sh}) that for 
\psd\ symbols, $\Op(b): {\cal S'} \to{\cal S'}$. Notice also
that, as usual, (\ref{Opb2}) implies that if $b$ depends
only on one canonical variable, for instance $p$, then $\Op(b) = b(P)$
in the spectral-theoretic sense. In particular, if the quantum Hamiltonian 
is $H_{m} := -(\frac{1}{8\pi^{2}}) \Delta_{x} = \Op(\Hm)$, we have
$\Op(e^{-\beta\Hm}) = e^{-\beta H_{m}}$.
\par
We can now calculate the {\em Weyl composition} $a \c b$ of two 
symbols $a$ and $b$, i.e. the (unique) symbol such that $\Op(a \c 
b) = \Op(a)\Op(b)$. This is done by means of (\ref{Opb1}), remembering 
that $T(\eta,\xi)$ obeys the multiplication law of the Heisenberg 
group and that a symbol is obtained by its correspondent operator by 
substituting $\ei{(p\cdot\eta + q\cdot\xi)}$ to $T(\eta,\xi)$ when 
the operator is in the form (\ref{Opb1}) -- compare (\ref{Fou-antitransf})
to (\ref{Opb1}). The result, after some manipulations of the 
integrals, is -- not surprisingly -- an adaptment of the 
corresponding formula for the Euclidean space case (\cite{fo}, \S 
2.1):
\begin{eqnarray}
	&\pha& (a \c b)(p,q)\b = \label{Weyl-comp} \\
	&=& \! \frac{1}{L^{2m}} \! \sum_{\xi_{1},\xi_{2}}
	\! \int \! \hat{a}(\eta_{1},\xi_{1}) 
	\hat{b}(\eta_{2},\xi_{2}) e^{\pi i(\eta_{1}\cdot\xi_{2} - 
	\xi_{1}\cdot\eta_{2})} \ei{[(\eta_{1}+\eta_{2})\cdot p + 
	(\xi_{1}+\xi_{2})\cdot q]} d\eta_{1} d\eta_{2} = \nonumber \\
	&=& \! \frac{1}{L^{2m}} \! \sum_{\xi_{1},\xi_{2}} \! \int \! 
	a \left( p \!+\! \frac{\xi_{1}}{2},q \!+\! q_{1} \right) 
	b \left( p \!+\! \frac{\xi_{2}}{2},q \!+\! q_{2} \right)
	\ei{(\xi_{2}\cdot q_{1} - \xi_{1}\cdot q_{2})} dq_{1}dq_{2}.
	\nonumber 
\end{eqnarray}
Notice that the sum is carried over $\xi_{1},\xi_{2} \in 
\TLms$ and the integration over $q_{1},q_{2} \in \TLm$.
\par
By the above formula we deduce that the Weyl composition has a 
property that may be called the {\em quasi-tracial property}:
\begin{equation}
	\int_{\LLm} (a \c b)(p,q) \, dpdq = \int_{\LLm} a(p,q) b(p,q) \, 
	dpdq,
	\label{quasi-tracial}
\end{equation}
even though we cannot hope to have the tracial property for some $\Ltwok$, 
i.e. (\ref{quasi-tracial}) with ${\rm Tr}_{\Ltwok} \Op(a)\Op(b)$ on 
the l.h.s., as explained in \cite{dbg}. Actually, as it will be clear 
in \S\ref{subs-Gibbs}, what appears in the l.h.s. is something 
resembling ${\rm Tr}_{L^{2}(\R^{m})}$.
\par 
Given $f^{(k)},g^{(k)} \in\Ltwok$, we define the \Fou-Wigner \fn\ 
relative to those two vectors as $V_{f^{(k)},g^{(k)}} (\eta,\xi) := 
\la f^{(k)},T(\eta,\xi) g^{(k)} \ra$. This is completely 
analogous to what is found in \cite{fo}. The Wigner \fn\ 
$W_{f^{(k)},g^{(k)}}$ is defined to be the (possibly distributional) 
\Fou\ transform of $V_{f^{(k)},g^{(k)}}$ and thus, from (\ref{Opb1}), 
\begin{equation}
	\la f^{(k)},\Op(b) g^{(k)} \ra_{\Ltwok} = \int_{\LLm} b(p,q) \,
	W_{f^{(k)},g^{(k)}}(p,q) \, dpdq;
\end{equation}
to be understood as $W_{f^{(k)},g^{(k)}}$ being the distribution
kernel of $b \mapsto \la f^{(k)},\Op(b) g^{(k)} \ra$. Again the
standard form for this ``\fn'' (\cite{fo}), calculated from (\ref{Opb2}) 
or from its very definition,
\begin{equation}
	W_{f^{(k)},g^{(k)}}(p,q) := \int_{\R^{m}} \emi{p\cdot z} 
	\overline{f^{(k)} (q-z/2)} \, g^{(k)}(q+z/2) \, dz.
	\label{Wigner-fn}
\end{equation}
has to interpreted in the weak sense.

\section{Statement of the Results} \label{sec-statement}
\sectcount

Suppose we have, $\forall L>0; m\in\Z, m\ge 1$ a measure space 
$(X_{L}^{m},d\theta)$
\footnote{We are intentionally a little informal here, in 
order to keep the notation not to cumbersome.  A better label for 
$d\theta$ would be $d\theta_{L}^{m}$.  The same for $d\nu$ in the 
following.} 
and a family of states
\begin{equation}
	\{ f_{\lambda} \}_{\lambda \in X_{L}^{m}} \subset \bigcup_{k \in 
	[0,1/L)^{m}} L_{(k)}^{2} (\TLm)
\end{equation}
labeled by the index $\lambda$ ranging in $X_{L}^{m}$. Call 
$w_{\lambda}(p,q) := W_{f_{\lambda},f_{\lambda}}(p,q)$, the Wigner 
\fn\ corresponding to $f_{\lambda}$.
\par
{\sc Hypothesis.} We suppose that
\begin{equation} 
	\int_{X_{L}^{m}} w_{\lambda}(p,q) d\theta(\lambda) \equiv 1,
	\label{hyp}
\end{equation}
as a distribution on $\LLm$.
\par 
{\sc Remark.} Such family of states represent in this context the 
quantum substitute for the classical phase space. As a matter of 
fact, (\ref{hyp}) says that the $\{ f_{\lambda} \}$ are evenly 
distributed, as a whole, over $\LLm$. Hence they play the role 
of ``points''. This is clearly seen in the case of the {\em \cs s}, 
perhaps the most remarkable example of states fulfilling the above 
hypothesis. They are introduced in \S\ref{appendix-cs} of the appendix.
However, (\ref{hyp}) can be restated by saying that we are given a set 
of states complete over all $\Ltwok, \, k\in [0,1/L)^{m}$ -- see 
\S\ref{subs-Gibbs}.
\par\smallskip
Since $\| e^{-\beta H_{m}} f_{\lambda} \|^{2} = \la f_{\lambda}, e^{-2\beta 
H_{m}} \, f_{\lambda} \ra$,
\footnote{Again a remark about the notation. A more formal symbol 
for the scalar product would be $\la \cdot, \cdot \ra_{\Ltwok}$
where $k=k(\lambda)$ is the uniquely determined $k \in [0,1/L)^{m}$ 
such that the arguments are in $\Ltwok$. Since the scalar 
products have the same structure on each $\Ltwok$ we will drop that 
subscript.}
then (\ref{hyp}) immediately implies
\begin{equation}
	\int_{X_{L}^{m}} \| e^{-\beta H_{m}} f_{\lambda} \|^{2} 
	d\theta(\lambda) = \int_{\LLm} e^{-2\beta\Hm(p)}\, dpdq = L^{m} \left( 
	\frac{\pi}{\beta} \right)^{m/2}.
	\label{Gibbs}
\end{equation}
We define:
\begin{equation}
	d\nu(\lambda) := \frac{\| e^{-\beta H_{m}} f_{\lambda} \|^{2} 
	d\theta(\lambda)} {\int_{X_{L}^{m}} \| e^{-\beta H_{m}} 
	f_{\lambda'} \|^{2} d\theta(\lambda')} 
	\label{dnu}
\end{equation}
and
\begin{equation}
	g_{\lambda} := \frac{e^{-\beta H_{m}} f_{\lambda}} {\|e^{-\beta H_{m}} 
	f_{\lambda} \|} 
\end{equation}
be the image under the quantum Gibbs measure of each of our states.
\begin{defn}
	$a\in {\cal A}$ is said an {\em asymptotic symbol} if $\exists 
	m_{0} \in\N, L_{0} >0$ such that $\forall m \ge m_{0}, L \ge L_{0}$, 
	$a\circ\PLm$ is a \psd\ symbol over $\LLm$.
	\label{def-asymp-sym}
\end{defn}
\par 
{\sc Remark.} Notice that with such a definition, asymptotic symbols 
are rather rigid objects. In fact, fixed an $m \ge m_{0}$, take
$L_{0} \le L \le L_{1}$. Now $\mbox{Im}(\PLm) \subseteq 
\mbox{Im}(\Pi_{L_{1}}^{m}) \subseteq \Linf$. So $a\circ\PLm$ is just 
the restriction of $a\circ \Pi_{L_{1}}^{m}$ to $\LLm$. But also 
$a\circ\PLm$, in order to be a symbol must be $C^{\infty}$ and 
$\TLm$-periodic. This means that $\forall i=1, \ldots, m$,
\begin{eqnarray}
	(a\circ \Pi_{L_{1}}^{m})(\ldots,-L/2,\ldots) &=&
	(a\circ \Pi_{L_{1}}^{m})(\ldots,+L/2,\ldots);  \\
 	\frac{d}{dq_{i}} (a\circ \Pi_{L_{1}}^{m})(\ldots,-L/2,\ldots) &=&
 	\frac{d}{dq_{i}} (a\circ \Pi_{L_{1}}^{m})(\ldots,+L/2,\ldots), 
\end{eqnarray}
where $(\ldots,\pm L/2,\ldots)$ stands for $(p_{1},\ldots,p_{m},q_{1}, 
\ldots,q_{i-1},\pm L/2, q_{i+1}, \ldots,q_{m})$.
Thus, parity arguments imply that for every $|q_{i}| \ge L_{0}/2$
\begin{equation}
	\frac{d}{dq_{i}} (a\circ \Pi_{L_{1}}^{m})(\ldots,q_{i},\ldots) = 0.
\end{equation}
Hence, for a fixed large $m$, $(a\circ \Pi_{L_{1}}^{m})$ is a 
constantwise continuation of $(a\circ \Pi_{L_{0}}^{m})$, and the 
former is completely determined by the latter. This also explain we 
had to ask for {\em asymptotic} symbols: one could not request 
the above property to hold $\forall m \ge 0, L \ge L_{0}$.
Examples of such \fn s are to be found in the families ${\cal 
B}^{(n)}$ defined after Lemma \ref{lemma1-mix} in \S\ref{sec-proofs}.
\smallskip\par 
We are now ready to state the theorems. For any operator $A$ acting over 
$L^{2}(\R^{m})$, define:
\begin{eqnarray}
	\Theta^{m}[t]\, (A) &:=& \ei{t H_{m}} A \emi{t H_{m}};
	\label{Theta-t}  \\
	\Xi^{m}[T]\, (A) &:=& \frac{1}{2T} \int_{-T}^{T} \Theta^{m}[t] (A) \, 
	dt. \label{Xi-T}
\end{eqnarray}
The quantum \mix\ property reads:
\begin{thm}
	Suppose $a,b$ asymptotic symbols in $L^{2}(\Linf,P_{\rho,2\beta})$ 
	and denote
	\begin{displaymath}
	 	I(t,L,m) := \int_{X_{L}^{m}}\la g_{\lambda}, \Theta^{m}[t]
 		(\Op(a\circ\PLm)) \, \Op(b\circ\PLm) \, g_{\lambda} \ra \, 
 		d\nu(\lambda).
	\end{displaymath}
	Then:
	\begin{displaymath}
		\lim_{|t|\to\infty} \tdlim I(t,L,m) = P_{\rho,2\beta}(a) 
		P_{\rho,2\beta}(b). 
	\end{displaymath}
	\label{thm-mix}
\end{thm}
\par
As regards the quantum \erg ity,
\begin{thm}
	Let $a$ be an asymptotic symbol in
	$L^{2}(\Linf,P_{\rho,2\beta})$. Let
	\begin{displaymath}
	 	J(T,L,m) := \int_{X_{L}^{m}} \| ( \Xi^{m}[T] 
	 	(\Op(a\circ\PLm)) - P_{\rho,2\beta}(a) ) \, g_{\lambda} 
	 	\|^{2}\, d\nu(\lambda).
	\end{displaymath}
	Then, for all $L,m$, the operator
	\begin{displaymath}
		\Xi^{m}[\infty] (\Op(a\circ\PLm)) := \lim_{T\to\infty} \Xi^{m}[T] 
		(\Op(a\circ\PLm))
	\end{displaymath}
	exists in the domain of $\Op(a\circ\PLm)$ and 
	\begin{displaymath}
		\tdlim J(\infty,L,m) = 0.
	\end{displaymath}
	Furthermore, if $a$ is also bounded, then the two limits can be 
	inverted:
	\begin{displaymath}
		\lim_{T\to\infty} \tdlim J(T,L,m) = 0.
	\end{displaymath}
	\label{thm-erg}
\end{thm}
\medskip\par
{\sc Remark.} The interchange of the time limit with the \td, in the  
above theorem, is remarkable. The fact that the time average can
be taken before the \td\ can be described saying that the  
finite-particle system (the ``real'' one) is {\em quasi-ergodic}, for very
large $L$ and $m$; that is, the time average of any {\em decent} 
\fn\ is close, in measure, to a constant. This is a feature of the 
classical ideal gas which has nothing to do with quantum mechanics.  
It is rather a consequence of the kinematic chaos and the restriction 
to symmetric observables, as anticipated in \S\ref{sec-intro}, remark (i). 
This can be seen quite easily, due to the integrability of the 
motion: time averaging means almost everywhere averaging over a torus.
The invariant tori here are the sets $\{ p \} \times \TLm \in \LLm$,
and so the time average of $a(p,q)$ is simply $\int a(p,q) dq$; the 
invariant \fn s would depend on $p$ only. Those \fn s, however, are 
requested to be symmetric and thus they cannot concentrate around a 
torus if they do not concentrate around all ``symmetric tori'' as well. 
An example is 
\begin{equation}
	a(p_{1},\, \ldots\, ,p_{m}) = \left\{
	\begin{array}{ll}
		1 & \mbox{if } p_{i} \in \Gamma \ \forall i=1,\ldots,m;  \\
		0 & \mbox{otherwise},
	\end{array}
	\right.
\end{equation}
where $\Gamma$ is a Borel set of $\R$. Now the kinematic chaos effect 
comes: at the \td, the support of this \fn, which is the probability 
to find all the particles having momenta in $\Gamma$, is 
exponentially small. 
\par
We can compare this to the situation one has for the harmonic chain, as 
shown in the companion paper \cite{gm}. In that case, there is no requirement 
on the observables. The fact that they cannot concentrate over 
invariant tori is instead due to the assumptions on the coupling matrix, 
which shuffles the tori at the infinite-particle limit.
\medskip\par
An even clearer reason for referring to Theorem \ref{thm-erg} as quantum 
\erg ity comes from the following:
\begin{cor}
	Assume $a$ bounded asymptotic symbol. Then, $\forall \epsilon>0$,
	set
	\begin{displaymath}
		K(\epsilon,T,L,m) := \nu \left( \{\lambda\in X_{L}^{m} \:\a\: 
		\left| \la g_{\lambda}, \Xi^{m}[T] (\Op(a\circ\PLm))\, g_{\lambda} 
		\ra - P_{\rho,2\beta}(a) \right| > \epsilon \} \right).
	\end{displaymath}
	Then
	\begin{displaymath}
		\lim_{T\to\infty} \tdlim K(\epsilon,T,L,m) =
		\tdlim \lim_{T\to\infty} K(\epsilon,T,L,m) = 0.
	\end{displaymath}
	\label{cor-erg}
\end{cor}
This can be phrased as follows. Call {\em $(T,\epsilon)$-exceptional 
initial states} those states $g_{\lambda}$ for which the quantum 
expectation of the $T$-time average is greater than $\epsilon$. Then the 
claim is that the measure of the $(T,\epsilon)$-exceptional initial 
states vanishes when the \td\ and the time limit are achieved.
\par
{\sc Proof of Corollary \ref{cor-erg}.} Easy consequence of Theorem 
\ref{thm-erg}, using a Cauchy-Schwartz inequality.
\qed

\subsection{The Quantum Gibbs State} \label{subs-Gibbs}

The results just formulated can be given a compact form, within the 
realm of the $C^{*}$ dynamical systems theory. It is beyond the 
purpose of this paper to go deep into that, so we do not outline the 
main notions of such a theory, hoping that the statements in this 
section are self-explanatory. However, a brief survey is given in 
\cite{gm}, Appendix 2. Here we just observe that the relations we will 
write are included in such a general frame. The interested reader is 
referred to \cite{br} for complete details, and to \cite{b} for a 
recent well-organized review.
\par
Consider ${\cal L}_{L}^{m}$, the space of all operators on 
$L^{2}(\R^{m})$ which are invariant and bounded over all the fibers 
$\Ltwok$. This is a $C^{*}$-algebra when endowed with the usual operator 
norm. Associated to this algebra we define the Heisenberg dynamics 
given by $\Theta^{m}[t]\, (A)$ as in (\ref{Theta-t}), and the quantum 
Gibbs state expressed by
\begin{equation}
	\omega_{\beta,L}^{m}(A) := \frac{\int_{[0,1/L)^{m}} \Tr (A \, 
	e^{-\beta H_{m}})\, dk}{\int_{[0,1/L)^{m}} \Tr (e^{-\beta H_{m}})
	\, dk} ,
	\label{omega}
\end{equation}
where $\Tr$ denotes the trace over $\Ltwok$. This functional is 
clearly normalized
\footnote{Here normalized means $\omega_{\beta,L}^{m}(\Id) = 1$.}
and invariant for the $^{*}$-automorphism $\Theta^{m}[t]$.  
Actually, it turns out to be a KMS state with parameter $\beta$ over 
the $W^{*}$ dynamical system $({\cal L}_{L}^{m},\Theta^{m}[t],
\omega_{\beta,L}^{m})$.
\par 
Now consider, for each value of $k \in [0,1/L)^{m}$, the standard 
\Fou\ basis $\{ e_{\alpha}^{(k)} \}$ as defined in  
\S\ref{subs-classical}. Next consider the family of all such vectors, 
labeled by the index $\lambda := (\alpha,k) \in X_{L}^{m} =: \TLms
\times [0,1/L)^{m}$. Endow $X_{L}^{m}$ with the measure
\begin{equation}
	d\theta(\alpha,k) := L^{m} \sum_{\xi\in\TLms} \delta(\alpha-\xi) 
	\, d\alpha\, dk.
\end{equation}
Such a family satisfies hypothesis (\ref{hyp}). In fact, calling 
$w_{\alpha,k}$ the Wigner function relative to $e_{\alpha}^{(k)}$, a 
straightforward computation from (\ref{Wigner-fn}) yields
\begin{equation}
	w_{\alpha,k}(p,q) = \frac{1}{L^{m}} \: \delta(p-(\alpha + k)).
	\label{w-alpha-k}
\end{equation}
Integrating this in $d\theta(\alpha,k)$ we obtain (\ref{hyp}). Thus, 
define, as it is done in at the beginning of this section, $g_{\alpha,k}:= 
e^{-\beta H_{m}} e_{\alpha}^{(k)} / \| e^{-\beta H_{m}} e_{\alpha}^{(k)} \|$. 
By definition (\ref{omega}), if $A \in {\cal L}_{L}^{m}$, we have
\footnote{Notice that we are using $2\beta$ instead of $\beta$ in the 
remainder.}
\begin{eqnarray}
	\omega_{2\beta,L}^{m}(A) &=& \frac{\int dk \sum_{\alpha} 
	\la e^{-\beta H_{m}} e_{\alpha}^{(k)}, \, A\,e^{-\beta H_{m}} 
	e_{\alpha}^{(k)} \ra} {\int dk \sum_{\alpha} \| e^{-\beta H_{m}}
	e_{\alpha}^{(k)} \|^{2} } = \nonumber \\
	&=& \int_{X_{L}^{m}} \la g_{\alpha,k}, \, A\, g_{\alpha,k} \ra \: 
	d\nu(\alpha,k),
\end{eqnarray}
with $d\nu$ defined as in (\ref{dnu}). On the other hand, if one calls in 
a natural way $A_{L}^{m} := \Op(a\circ\PLm)$, then a simple argument which 
is better explained in the following (see formul\ae\ (\ref{mix-zero}) 
and (\ref{mix-first})) gives
\begin{equation}
	\omega_{2\beta,L}^{m}(A_{L}^{m}) = \int_{\LLm} (a\circ\PLm) (p,q) 
	\left[ e^{-2\beta\Hm(p)}\, dpdq \right]_{N}.
\end{equation}
Proposition \ref{limit-of-meas} immediately yields
\begin{equation}
	\tdlim \omega_{2\beta,L}^{m}(A_{L}^{m}) = P_{\rho,2\beta}(a).
\end{equation}
Hence, if $a,b$ are asymptotic symbols, we have just proved that Theorem 
\ref{thm-mix} can be rewritten as
\begin{equation}
	\lim_{|t|\to\infty} \tdlim \omega_{2\beta,L}^{m} \left( 
	\Theta^{m}[t] (A_{L}^{m}) \, B_{L}^{m} \right) = 
	\tdlim \omega_{2\beta,L}^{m}(A_{L}^{m}) 
	\cdot \tdlim \omega_{2\beta,L}^{m}(B_{L}^{m}).
	\label{thm-mix-1}
\end{equation}
In the same spirit, Theorem \ref{thm-erg} becomes
\begin{eqnarray}
	&\pha& \lim_{T\to\infty} \tdlim \omega_{2\beta,L}^{m} ( (\Xi^{m}[T] 
	(A_{L}^{m}))^{2}) =  \nonumber \\
	&=& \tdlim \omega_{2\beta,L}^{m} ( (\Xi^{m}[\infty] 
	(A_{L}^{m}))^{2}) =  \label{thm-erg-1} \\
	&=& \tdlim ( \omega_{2\beta,L}^{m} (A_{L}^{m}) )^{2},
	\nonumber 
\end{eqnarray}
\par
{\sc Remark.} We have not defined an algebra of quantum observables for 
the infinite-particle system, limiting ourselves to deal with finite
dimensions and to take a \td\ afterwards (see also comment 3 below). Had 
we introduced such a mathematical framework, then relations
(\ref{thm-mix-1}) and (\ref{thm-erg-1}), for the state 
$\tdlim \omega_{2\beta,L}^{m}$, would be contained in the general set of 
chaoticity notions in $C^{*}$ dynamical system theory (see \cite{b}, 
Definitions 4.42, 4.43).
\footnote{We are not granted, in principle, all the equivalent \erg ity 
and \mix\ properties recalled in that reference, since we have not 
proved asymptotic commutativity.} 
\par
Some comments concerning the above reformulation of the theorems as 
compared to \cite{gm}:
\begin{enumerate}
\item 
(\ref{thm-mix-1}) is completely analogous to statement (1.10) in 
\cite{gm}. That is, the quantum \mix, forbidden in the finite-particle 
frame by the quasi-periodicity of the Heisenberg evolution, regardless the 
dynamics of the classical flow (see \S\ref{sec-intro}), is restored at 
the \td.
\item  
A formulation of the \erg ity similar to (1.9) in \cite{gm} has not 
been chosen in this context because of the technicalities it would 
require.  The \cs s we have here (see appendix, \S\ref{appendix-cs}) 
are indexed by $\lambda \in \TLm \times \R^{m} \times [0,1/L)^{m} =: 
X_{L}^{m}$, which is not exactly the classical phase space $\LLm$.  
However, one could explicitly calculate $d\nu$ over $X_{L}^{m}$, for 
particular choices of the \cs s, and find a limit measure space -- say 
-- $(X,d\nu)$, but this turns out to be rather cumbersome and possibly 
misleading. Understandably, though, (\ref{thm-erg-1}) and especially 
Corollary \ref{cor-erg} carry the same physical meaning as the mentioned 
result.
\item
As already exploited, we are able here to state the \erg ity results 
with a commutation of the limits.
\item
As emphasized in \cite{gm}, \S1, Remark 2, the techniques we use 
to prove the quantum \erg\ properties at the \td, have 
the useful outcome to show that the r.h.sides of (\ref{thm-erg-1}) and
(\ref{thm-mix-1}) are the expected classical Gibbs averages. This 
is why we have formulated Theorems \ref{thm-erg} and \ref{thm-mix} 
in the first place. 
\item
More importantly, here, results (\ref{thm-mix-1}) and (\ref{thm-erg-1}) 
were not known, at least to us.
\end{enumerate}

\section{The Proofs.} \label{sec-proofs}
\sectcount

The first key fact is the following
\begin{lemma} \label{lemma-q-evol-c}
	For every symbol $c$ defined on $\LLm$,
	\begin{displaymath}
		\ei{t H_{m}} \Op(c) \emi{t H_{m}} = \Op(c\circ\fLmt)
	\end{displaymath}
\end{lemma}
This is true since we are dealing with a linear flow.  This property 
of linear flows -- sometimes referred to as the {\em exact Egorov 
Theorem} for the evolution canonical transformation -- dates back at 
least to Van Hove and is valid only for Weyl quantization, whose 
restriction to $L$-periodic symbols we are 
now using.  Anyway, for the sake of  
convenience, a direct proof is found in 
\S\ref{appendix-pf-quant-evol} of the appendix.
\smallskip\par
In our case, applying this lemma to $a \circ \PLm$ and using the 
remark in formula (\ref{comm-of-flows}), we have
\begin{equation}
	\ei{t H_{m}} \Op(a\circ \PLm) \emi{t H_{m}} = \Op(a \circ \finft 
	\circ \PLm).
	\label{quant-evol-comm}
\end{equation}

\subsection{Proof of Theorem 3.2.} \label{subs-pf-thm-mix}

In view of the above relation we call $a_{t} := a \circ \finft$.  In 
the rest of this proof, whenever there is no confusion, we denote 
by a quote the immersion application from $\LLm$ to $\Linf$. Hence $a' 
:= a \circ \PLm$ and so on. In other words, $a'$ is just our 
observable $a$ looked at in the finite dimensional 
phase space $\LLm$. By (\ref{hyp}) and (\ref{dnu}), the
definition of $I$, in the statement of the theorem, yields
\begin{equation}
	I(t,L,m) = \frac{\int_{\LLm} (e^{-\beta\Hm} \c a'_{t} \c 
	b' \c e^{-\beta\Hm}) dpdq} {\int_{\LLm} e^{-2\beta\Hm} 
	dpdq}.
	\label{mix-zero}
\end{equation}
Using twice (\ref{quasi-tracial}) -- once to permutate cyclically
the factors in (\ref{mix-zero}) and once to remove one of the $\c$ 
signs -- leads to
\begin{equation}
	I(t,L,m) = \int_{\LLm} (a'_{t} \c b') (p,q) 
	\left[ e^{-2\beta\Hm(p)}\, dpdq \right]_{N},
	\label{mix-first}
\end{equation}
since obviously $e^{-\beta\Hm} \c e^{-\beta\Hm} = e^{-2\beta\Hm}$.
We further denote by $\mLm$ the classical 
Gibbs measure (at inverse temperature $2\beta$) over $\LLm$: 
$\mLm(p,q) := L^{-m}(\pi/\beta)^{-m/2} e^{-\beta p^{2}}$, and by 
$\mbm$ its component in the $p$-space: $\mbm(p) := (\pi/\beta)^{-m/2} 
e^{-\beta p^{2}}$.
\par
In view of (\ref{Weyl-comp}), (\ref{mix-first}) becomes, after 
some elementary but tedious rearrangements of the nested integrals,
\begin{equation}
	I(t,L,m) = \int_{\LLm} [a'_{t}\, (b' *_{L} \Phi_{L}^{m})] 
	(p,q) \: d\mLm(p,q),
	\label{mix-second}
\end{equation}
where 
\begin{equation}
	\Phi_{L}^{m}(p,q) = e^{\beta p^{2}} \frac{1}{L^{m}} 
	\sum_{\xi\in\TLms} e^{-\beta(p-\xi/2)^{2}} \, \ei{\xi\cdot q},
	\label{phiLm-1}
\end{equation} 
and $*_{L}$ means convolution in the $q$-variable on $\TLm$. A 
particular care must be taken here about this convolution {\em on a 
torus}, in order to prevent mistakes: see \S\ref{appendix-conv}.
\par
Thus $\Phi_{L}^{m}$ is completely factorizable, with $\Phi_{L}$ being
a natural symbol for each of his factor: if 
$f(p_{1}, \ldots,p_{m}, q_{1}, \ldots, q_{m}) 
= f_{1}(p_{1}, q_{1}) \cdots f_{m}(p_{m}, q_{m})$, then
\begin{equation}
	(f *_{L} \Phi_{L}^{m})(p,q) = (f_{1} *_{L} \Phi_{L})(p_{1},q_{1})
	\cdots (f_{m} *_{L} \Phi_{L})(p_{m},q_{m}).
	\label{phiLm-2}
\end{equation}
This property will be useful in the following.
Also, if $1$ is the \fn\ on $\LLm$ identically equal to 1,
\begin{equation}
	1 *_{L} \Phi_{L}^{m} = 1.
	\label{phiLm-3}
\end{equation}
We are going to prove the statement of the theorem, starting from 
(\ref{mix-second}), for $b$ in a dense subspace of 
$L^{2}(\Linf,P_{\rho,2\beta})$. To accomplish that we need to make 
the following construction.
\par
Fix a positive integer $n$ and consider the \fn\ $\beta(p,q) \in 
C_{0}^{\infty}(\R^{2n})$  
\footnote{Not to be confused with the inverse 
temperature $\beta$, a fixed parameter throughout this paper.}
, that is, infinitely differentiable \fn s with compact 
support. For $m > n$ define the application
\begin{equation}
	\NLmn(\beta)(p_{1}, \ldots, p_{m}, q_{1}, \ldots, q_{m}) := 
	\sum_{j_{1}=1}^{m} \cdots \sum_{j_{n}=1}^{m} \beta (p_{j_{1}}, 
	\ldots, p_{j_{n}}, q_{j_{1}}, \ldots, q_{j_{n}}).
	\label{NLm}
\end{equation}
So $\NLmn(\beta)$ is a \fn\ defined on $\R^{2m}$, and thus in particular
on $\LLm$. The use of this application is explained by next 
\begin{lemma} \label{lemma1-mix}
	If $\beta(p,q) \in {\cal S} (\R^{2n})$ 
\footnote{This denotes the Schwartz class.}
	then there exists a \fn\ $b\in L^{2}
	(\Linf,P_{\rho,2\beta})$ s.t. $\NLmn(\beta) = b \circ \PLm$.
	Plus, the following properties hold:
	\begin{equation}
		\int_{\LLm} |(b\circ\PLm) (p,q)|^{2} \: d\mLm(p,q) 
		\le m^{2n} \int_{\R^{2n}} |\beta(p',q')|^{2} \: d\mu_{L}^{n}
		(p',q') ;
		\label{lemma1-one} 
	\end{equation}
	\begin{equation}
		\int_{\LLm} (b\circ\PLm) (p,q) \: d\mLm(p,q) =  
		\frac{m!}{L^{m} (m-n)!}  \int_{\R^{2n}} \beta(p',q') 
		\: d\breve{\mu}^{n} (p')\, dq'. 
		\label{lemma1-two}
	\end{equation}
\end{lemma}
\par
{\sc Proof of Lemma \ref{lemma1-mix}.} The first inequality simply 
follows by definition (\ref{NLm}): we have $m^{2n}$ integrals over 
$\LLm$. To the cross-term integrals we apply Cauchy-Schwartz in order 
to obtain $m^{2n}$ terms equal to $\| \beta \|^{2}_{L^{2}(\LLm)}$.  
These get reduced to integrals over $\LLn$, since the 
measure is decomposable and the integrand \fn s depend only on $2n$ 
variables; finally they are extended to all of $\R^{2n}$. 
\par 
To prove the rest we approximate $\beta$ with suitably 
chosen indicator \fn s over $\R^{2n}$.  More precisely take two 
sufficiently fine partitions of $\R$, $\{ \Gamma_{j} \}, \{ 
\Delta_{\ell} \}\subset {\cal B}(\R)$ with $\sup_{j,\ell} \{ 
\mbone(\beta_{j}), |\Delta_{\ell}| \}$ small.
\footnote{$| \cdot |$ denotes the Lebesgue measure.}
Let $\chi_{j_{1},\ldots,j_{n}, \ell_{1},\ldots,\ell_{n}} 
(p',q')$ be the indicator \fn\ of the set $\Gamma_{j_{1}} \times \cdots 
\times \Gamma_{j_{n}} \times \Delta_{\ell_{1}} \times \cdots \times
\Delta_{\ell_{1}}$ and approximate $\beta$ with $\beta_{a} := 
\sum c_{j_{1},\ldots,\ell_{n}}\, \chi_{j_{1},\ldots,\ell_{n}}$. So
\begin{equation}
	\NLmn(\beta_{a})(p,q) = \sum_{j_{1},\ldots,\ell_{n}} 
	c_{j_{1},\ldots,\ell_{n}} \NLmn(\chi_{j_{1},\ldots,\ell_{n}})(p,q).
	\label{step-in-lemma1}
\end{equation}
Since $\chi_{j_{1},\ldots,\ell_{n}}$ is completely factorizable, 
then it is easy to realize, by definition (\ref{NLm}), 
that $\NLmn(\chi_{j_{1},\ldots,\ell_{n}}) (p,q) = N_{L,m}^{(1)} 
(\chi_{\Gamma_{j_{1}} \times \Delta_{\ell_{1}}}) \cdots N_{L,m}^{(1)} 
(\chi_{\Gamma_{j_{n}} \times \Delta_{\ell_{n}}})$. We see that 
$N_{L,m}^{(1)}(\chi_{\Gamma \times \Delta}) (p,q)$ takes integer 
values between $0$ and $m$. Specifically if counts the number of 
particles in the configuration $(p,q) \in\LLm$ whose momentum is 
contained in $\Gamma$ and whose coordinate in $\Delta$.
\footnote{This explains why we have chosen such a notation for 
$\NLmn$.}
So $N_{L,m}^{(1)}(\chi_{\Gamma \times \Delta}) = N_{\Gamma \times 
\Delta} \circ \PLm$, where $N_{\Gamma \times \Delta}: \Linf \to \N$ 
is defined by
\begin{equation}
	N_{\Gamma \times \Delta} (p,q) := \# \{ x \in q \cap \Delta 
	\,|\, p(x) \in \Gamma \},
\end{equation}
where, with sloppy notation, $(p,q)$ denotes a point in $\Linf$.  
Recalling what we said in \S\ref{subs-classical}, this \fn\ 
obviously belongs to ${\cal A}$: see in particular 
(\ref{B-Delta,Gamma}) and comments thereby. Therefore so does every finite 
product of similar \fn s. Looking at (\ref{step-in-lemma1}), and 
subsequent comments, this proves that there exists a $b_{a} \in 
{\cal A}$ such that $\beta_{a} = b_{a}\circ\PLm$. The analogous 
statement holds for $\beta$ as well, by density.
\par
Let us go over to the proof of (\ref{lemma1-two}).  Fix a sequence 
$(j,\ell):=(j_{1},\ldots,\ell_{n})$ like those we have in formula 
(\ref{step-in-lemma1}) and fix $n$ integers $k := (k_{1}, \ldots, 
k_{n})$ such that $k_{1} + \ldots + k_{n} \le m$.  Now consider the 
set $A_{j,\ell}^{(k)} := \{ N_{L,m}^{(1)} (\chi_{\Gamma_{j_{i}} \times 
\Delta_{\ell_{i}}}) = k_{i} \, ;\, \forall i=1, \ldots, n \} \in \LLm$, 
i.e. the set of the configurations having $k_{1}$ particles in 
$\Gamma_{j_{1}} \times \Delta_{\ell_{1}}$, $k_{2}$ particles in 
$\Gamma_{j_{2}} \times \Delta_{\ell_{2}}$, and so on.  Notice that 
(use some combinatorics and, anyway, refer to \cite{s})
\begin{equation}
	\mLm( A_{j,\ell}^{(k)} ) = \frac{m!} {(m -\sum k_{i})!}
	\left[ \prod_{i=1}^{n} \frac{1}{k_{i}!} 
	\left( \frac{ |\Delta_{\ell_{i}}| }{L} \mbone(\Gamma_{j_{i}}) 
	\right)^{k_{i}} \right] \left( 1 - \sum_{i=1}^{n} 
	\frac{ |\Delta_{\ell_{i}}| }{L} \mbone(\Gamma_{j_{i}})
	\right)^{m -\sum k_{i}}.
\end{equation}
We have seen that $\NLmn(\chi_{j_{1},\ldots,\ell_{n}}) = 
\sum_{k} k_{1} \cdots k_{n} \, A_{j,\ell}^{(k)}$. So, when the 
initially chosen partition is fine, $\mLm(\chi_{j_{1},\ldots,
\ell_{n}}) = m! /(L^{m} (m-n)!) \prod_{i} |\Delta_{\ell_{i}}|
\mbone(\Gamma_{j_{i}}) + o(|\Delta_{\ell_{i}}|, \mbone(
\Gamma_{j_{i}}) )$. Looking back at (\ref{step-in-lemma1}) this proves 
that (\ref{lemma1-two}) holds with negligible errors for $\beta_{a}$ 
and thus is exact for $\beta$.
\qed
\par\medskip
Let us call ${\cal B}^{(n)} \in {\cal A}$ the space of \fn s $b$
granted by Lemma \ref{lemma1-mix} when $\beta \in C_{0}^{\infty}
(\R^{2n})$. From now on we will suppose $b\in {\cal B}^{(n)}$, 
so that $b' := b\circ\PLm = \NLmn(\beta)$.  In so doing we 
will be proving Theorem \ref{thm-mix} for $b \in \oplus_{finite} {\cal 
B}^{(n)}$.  But this is dense in $L^{2}(\Linf,P_{\rho,2\beta})$ since, 
looking at the proof of Lemma \ref{lemma1-mix}, the closure of 
${\cal B}^{(n)}$ contains the product of $n$ \fn s like $N_{\Gamma 
\times \Delta}$. This means that in the algebra $\overline{(\oplus\,
{\cal B}^{(n)})}$ we are able to find the indicator \fn s of the sets 
$N_{\Gamma \times \Delta}^{-1} (n),\, \forall n\in\N$. But these 
generate ${\cal A}$ (look at \ref{B-Delta,Gamma} and refer to 
\cite{gla} and \cite{b}).
\par
Under the above assumption, we go back to (\ref{mix-second}): since $b' = 
\NLmn(\beta)$, then $b' *_{L} \Phi_{L}^{m} = \NLmn(\beta *_{L} 
\Phi_{L}^{n})$ 
\footnote{Where this time $*_{L}$ means convolution in the 
$q$-variable over $\TLn$. We warn the reader again about the possible 
confusion arising from the fact that $\beta(p,q)$ is defined on $\R^{2n}$. 
When in a $*_{L}$-convolution, it has to be considered as restricted 
to $[-L/2,L/2)^{n}$ and periodic according to the identification 
$[-L/2,L/2)^{n} \simeq \TLn$. See again \S\ref{appendix-conv}.}
because of the mentioned properties of $\Phi_{L}^{m}$ 
(see (\ref{phiLm-1}) to (\ref{phiLm-3})). If we denote by 
$\gamma_{L} := \beta *_{L} \Phi_{L}^{n}$, it is obvious that 
$\gamma_{L} \in {\cal S}(\R^{2n})$ and so $\NLmn(\gamma_{L}) = c'_{L}$ 
for some $c_{L} \in L^{2}(\Linf,P_{\rho,2\beta})$, by Lemma 
\ref{lemma1-mix}. This allows us to rewrite (\ref{mix-second}) as
\begin{equation}
	I(t,L,m) = \la a'_{t}, c'_{L} \ra_{L^{2}(\LLm,\mLm)}.
	\label{mix-third}
\end{equation}
If we are able to find a limit for $c_{L}$ then we are done with the 
cumbersome part of this proof. To this goal, we formulate the following
\begin{lemma} \label{lemma-mix-tech}
	There exists a $\gamma_{\infty} \in {\cal S} (\R^{2n})$ such that
	\begin{equation}
		\| \gamma_{\infty} - \gamma_{L} \|^{2}_{L^{2} (\LLn, 
		\mu_{L}^{n})} = \OL.
		\label{lemma-t-one}
	\end{equation}
	Furthermore
	\begin{equation}
		\int_{\R^{2n}} \gamma_{\infty}(p',q') \, d\breve{\mu}^{n} 
		(p')\, dq' = \int_{\R^{2n}} \beta(p',q') \, d\breve{\mu}^{n}
		(p')\, dq'.
		\label{lemma-t-two}
	\end{equation}
\end{lemma}
The proof of this lemma is found in \S\ref{appendix-mix-tech}.
\medskip\par
In analogy with the above notations we call $c_{\infty}$ the 
observable in \linebreak[3] 
$L^{2}(\Linf,P_{\rho,2\beta})$ obtained applying Lemma 
\ref{lemma1-mix} to $\gamma_{\infty}$. Comparing now 
(\ref{lemma-t-one}) in Lemma \ref{lemma-mix-tech} with 
(\ref{lemma1-one}) we deduce that
\begin{equation}
	\tdlim \| c'_{\infty} - c'_{L} \|^{2}_{L^{2} (\LLm,\mLm)} = 0.
	\label{mix-fourth}
\end{equation}
Dropping for the sake of simplicity the subscript in the scalar 
product notation, this means that, when $m,L \to\infty,\, m/L \to\rho$,
\begin{eqnarray}
	&\pha& | \la a'_{t}, c'_{L} \ra - P_{\rho,2\beta}(a_{t} 
	c_{\infty}) | \le  \\
	&\le& \| a'_{t} \|^{2} \| c'_{L} - c'_{\infty} \|^{2} +
	| \mLm (a'_{t} c'_{\infty}) - P_{\rho,2\beta}(a_{t} c_{\infty}) |
	\to 0, \nonumber
\end{eqnarray}
because of Proposition \ref{limit-of-meas}. Now we use the other main 
ingredient of this proof, i.e. the classical result, Theorem \ref{vs-thm}. 
We obtain
\begin{equation}
	\lim_{|t|\to\infty} \tdlim I(t,L,m) = P_{\rho,2\beta}(a) 
	P_{\rho,2\beta}(c_{\infty}).
\end{equation}
Now, using the integrals of $\gamma_{\infty}$ and $\beta$ to compare 
the integrals of $c_{\infty}$ and $b$ (apply (\ref{lemma-t-two}) into 
(\ref{lemma1-two})), we see that $\mLm(c'_{\infty}) = \mLm(b')$. 
Taking the limits, $P_{\rho,2\beta}(c_{\infty}) = P_{\rho,2\beta}(b)$, 
which, together with the last relation, completes the proof.
\qed

\subsection{Proof of Theorem 3.3.} \label{subs-pf-thm-erg}

First of all it has to be noticed that both statements of Theorem 
\ref{thm-erg} (respectively relation (\ref{thm-erg-1})) cannot be 
derived so trivially from Theorem \ref{thm-mix} (resp.  
(\ref{thm-mix-1})). This will be seen below in each case. 
\par
We borrow the notation from the previous proof: so, for 
example, $a'_{t} := a \circ \finft \circ \PLm$.  Also, let $a'_{T} := 
(1/2T) \int_{-T}^{T} a'_{t} dt$. Formula (\ref{quant-evol-comm}) proves 
that
\begin{equation}
	\Xi^{m}[T] \, (A_{L}^{m}) = \Op(a'_{T}),
	\label{erg-00}
\end{equation}
where, as in \S\ref{subs-Gibbs}, we call $A_{L}^{m} := \Op(a')$.
\par
The existence of $\Xi^{m}[\infty] (A_{L}^{m})$ is a trivial 
consequence of the Heisenberg evolution: we can easily figure it out 
looking at its matrix elements w.r.t. the 
bases $\{ e_{\alpha}^{(k)} \} \subset \Ltwok$. These bases 
diagonalize the Hamiltonian $H_{m}$, as well as any operator \fn\ 
of $P$ only. We call such eigenvalues
\begin{equation}
	E_{\alpha}^{(k)} = \frac{1}{2} (\alpha + k)^{2} = \frac{1}{2}
	\sum_{i=1}^{n} (\alpha_{i} + k_{i})^{2}.
	\label{eigenv}
\end{equation}
Now it is easy to see that, $\forall\, k\in [0,1/L)^{m},\, 
\alpha,\gamma \in \TLms$,
\begin{equation}
	\la e_{\alpha}^{(k)},\, \Xi^{m}[\infty] (A_{L}^{m})\, e_{\gamma}^{(k)} 
	\ra = \la e_{\alpha}^{(k)},\, A_{L}^{m}\, e_{\gamma}^{(k)} \ra \:
	\delta_{E_{\alpha}^{(k)},E_{\gamma}^{(k)}},
	\label{erg-0}
\end{equation}
where $\delta$ is the Kronecker $\delta$-\fn. This formula
shows that $\Xi^{m}[\infty] (A_{L}^{m})$ is well defined on all vectors 
in $D(A_{L}^{m})$. 
\par
One might think to prove now the statement regarding $J(\infty,L,m)$ 
by simply substituting the Heisenberg invariant operator 
$\Xi^{m}[\infty] (A_{L}^{m})$ to $\Op(a')$ and $\Op(b')$ in Theorem 
\ref{thm-mix}. We cannot quite do this, since such operator is not 
in general \psd. It is obvious, though, that it can be 
approximated to any extent by \psd\ operators, and the result 
would follow by density. However, as remarked in \S\ref{subs-Gibbs}, 
we have not defined a proper $C^{*}$-algebra for the infinite-particle 
system. Thus, we cannot talk of any density and have to prove the 
theorem directly.
\par\noindent 
$\Xi^{m}[\infty] (A_{L}^{m})$, 
roughly speaking, represents the 
quantization of $\displaystyle a'_{\infty} :=
\lim_{T\to\infty} a'_{T}$, which  is not in
general a symbol, being possibly not even 
continuous.  But simple considerations based
upon the trivial dynamics  over $\LLm$ (see also
the remark after the statement of this theorem) 
show that it is almost everywhere (namely for
$p=(p_{1}, 
\ldots,p_{m})$ having rationally independent components) equal to
\begin{equation}
	c'(p,q) = c'(p) := \frac{1}{L^{m}} \int_{\TLm} 
	a'(p,q) \, dq
	\label{erg-c-prime}
\end{equation}
which is a symbol. Moreover we denote it $c'$
since one can  straightforwardly find a $c\in
{\cal A}$ such that $c' = c\circ\PLm$.  The
whole idea of this proof is exactly to show
that, in some  sense, $\Op(c')$ is a.e. equal to
$\Xi^{m}[\infty] (A_{L}^{m})$, so that the 
former can be substituted to the latter in the
definition of 
$J(\infty,L,m)$ in order to apply Theorem \ref{thm-mix} (also compare
(\ref{thm-erg-1}) and (\ref{thm-mix-1})). 
\par
We first remark some basic properties of $\Op(c')$.  Since $c'(p,q) = 
c'(p)$ then $\Op(c')$ is diagonal w.r.t.  $\{ e_{\alpha}^{(k)} \}$.  
Its diagonal matrix elements, using (\ref{w-alpha-k}), are found to be
\begin{equation}
	\la e_{\alpha}^{(k)},\, \Op(c')\, e_{\alpha}^{(k)} \ra =
	\frac{c'(\alpha+k)} {L^{m}} = \frac{1} {L^{m}} \int_{\TLm} 
	a'(\alpha+k,q) dq = \la e_{\alpha}^{(k)},\, A_{L}^{m}\, 
	e_{\alpha}^{(k)} \ra,
	\label{erg-1}
\end{equation}
showing incidentally, as it ought to be, that $\Op(c')$ is 
invariant for time evolution. More importantly, (\ref{erg-1}), 
together with (\ref{erg-0}), implies that $\Op(c')^{(k)} = 
(\Xi^{m}[\infty] (A_{L}^{m}))^{(k)}$
\footnote{Remember that with $A^{(k)}$ we denote $A_{|\Ltwok}$ as 
explained in \S\ref{subs-quant}.}
for those $k\in [0,1/L)^{m}$ for which $H_{m}^{(k)}$ is diagonal.
\par
Using (\ref{Wigner-fn}) over a generic $f_{\lambda}$ picked up from 
the set of states satisfying (\ref{hyp}), one can see that 
$w_{\lambda}(p,q)$ contains a (possibly countable) sum of $\delta$-\fn 
s in $p$.  This simple argument shows that, in order for $\{ 
f_{\lambda} \}$ to verify (\ref{hyp}), a factor of the measure space 
$(X_{L}^{m}, d\theta)$ must be $([0,1/L)^{m},d\tau(k))$, with $d\tau$ 
absolutely continuous w.r.t. the Lebesgue measure.
\footnote{This is a manifestation of the fact that all fibers 
$\Ltwok$ need to be taken into account, as mentioned in 
\S\ref{subs-quant}, Remark 1. We can convince ourselves of this also 
looking at the two examples of $\{ f_{\lambda} \}$ we have explicitly 
written: the \Fou\ basis in \S\ref{subs-Gibbs}
and the \cs s in the appendix,
\S\ref{appendix-cs}. In both cases $(X_{L}^{m},
d\theta) =  ([0,1/L)^{m},dk)) \times$ some
measure.} So if we prove that
$\sigma(H_{m}^{(k)})$ is simple for 
Lebesgue-almost all $k$'s, then
\begin{equation}
 	J(\infty,L,m) := \int_{X_{L}^{m}} \| (\Op(c') - P_{\rho,2\beta}(a) ) 
 	\, g_{\lambda} \|^{2}\, d\nu(\lambda)
\end{equation}
and we can apply Theorem \ref{thm-mix} with $a = b = c - P_{\rho,2\beta}(a)$, 
which is time invariant. This would complete the proof of the first 
claim.
\par
Rescaling (\ref{eigenv}) by a factor $L^{m}$, what we need is 
equivalent to the following
\begin{lemma}
	\begin{displaymath}
		\a \{ k\in [0,1)^{m} \,|\, \exists j,n \in \Z^{m}\: 
		s.t. \: (j+k)^{2} = (n+k)^{2} \} \a = 0.
	\end{displaymath}
	\label{lemma-lattice}
\end{lemma}
\par
{\sc Proof of Lemma \ref{lemma-lattice}.} 
\footnote{I thank D.Dolgopyat for this simple proof.}
Thinking of it as a geometric problem in $\R^{m}$, when such $j,n$ exist, 
then $-k$ lies in the axial hyperplane of the segment joining $j$ to 
$n$, i.e. the set of points in the space equally distant from $j$ and $n$. 
By very construction there is only a countable number of such hyperplanes.
\qed
\medskip\par
As far as the last statement of Theorem \ref{thm-erg} is concerned, 
we see again that it cannot be derived as a corollary of the mixing 
theorem since we are taking time limits of both operators. But we can 
give a direct proof using the techniques of \S\ref{subs-pf-thm-mix} 
and the classical \erg ity result contained in Theorem \ref{vs-thm}.
\par
Exactly as in (\ref{mix-zero}) and (\ref{mix-first}) we can write
\begin{equation}
	J(T,L,m) = \int_{\LLm} \left[ (a'_{T} - P_{\rho,2\beta}(a)) \c 
	(a'_{T} - P_{\rho,2\beta}(a)) \right] (p,q) 
	\left[ e^{-2\beta\Hm(p)}\, dpdq \right]_{N},
	\label{erg-2}
\end{equation}
having used (\ref{erg-00}). We have become familiar with this object 
in \S\ref{subs-pf-thm-mix}, and we have seen that integrating -- w.r.t. 
the Gibbs measure -- the Weyl composition of two \fn s means
integrating the product of the two \fn s, one of which scrambled by a 
convolution (see (\ref{mix-first}), (\ref{mix-second}) and 
(\ref{mix-third})). At the thermodynamic limit, this amounts to say 
that
\begin{equation}
	\tdlim J(T,L,m) = P_{\rho,2\beta} ((a_{T} - P_{\rho,2\beta}(a)) 
	\, c^{(T)}),
	\label{erg-3}
\end{equation}
where $c^{(T)}$ is the limit of the ``scrambled \fn s'' constructed 
upon $(a_{T} - P_{\rho,2\beta}(a))$. Its existence is granted by Lemmas 
\ref{lemma-mix-tech} and \ref{lemma1-mix}, which yielded 
(\ref{mix-fourth}). From the construction we have just recalled it 
can be seen that if $(a_{T} - P_{\rho,2\beta}(a))$ is bounded then 
$c^{(T)}$ is as well. 
\par
A remark is in order here: in \S\ref{subs-pf-thm-mix} we have worked 
with symbols belonging to ${\cal B}^{(n)}$, and those are unbounded 
by definition, being the limits of \fn s like $\NLmn(\beta)$ defined 
in (\ref{NLm}). But a simple argument shows that a bounded $a \in 
{\cal A} = \sigma (\oplus \, {\cal B}^{(n)})$ remains bounded after 
the above procedure, since, roughly speaking, it gets deformed in the 
same way in each of its ${\cal B}^{(n)}$-components. 
\par
Finally, we can apply Lebesgue dominated convergence in (\ref{erg-3})  
since the integrand \fn\ is bounded and tends pointwise 
to zero as $T\to\infty$. Thus, the $T$-limit of (\ref{erg-3}) gives 
the last statement in Theorem \ref{thm-erg}, whence the end of the proof.
\qed

\section{Acknowledgments}

I would like to thank S.Graffi and Ya.G.Sinai for addressing this 
problem to me, giving stimulation as well as useful advices. I also 
wish to thank A.Parmeggiani and A.Martinez for interesting 
discussions on this subject. A grant from I.N.F.N. is highly 
acknowledged.

\appendix
\section{Appendix}
\sectcount

\subsection{Coherent states for the cylinder.} \label{appendix-cs}

We begin this section by recalling some notions about the Bloch 
decomposition (\ref{Bloch-dec}), following \cite{rs} and \cite{dbg}.
The idea is very simple: given a \fn\ $f\in L^{2}(\R^{m})$, and 
therefore its \Fou\ transform $\hat{f}(p)$, we pick up from the 
latter only the terms at $p=\xi+k,\ (\xi\in\TLms)$ to construct 
$f^{(k)}$ which clearly lies in $\Ltwok$.
\par
In formula
\begin{equation}
	f^{(k)}(x) := \frac{1}{L^{m}} \sum_{\xi\in\TLms} \hat{f}(\xi+k)
	\ei{(\xi+k)\cdot x}.
	\label{def-fk}
\end{equation}
Considering the scalar products in the dual spaces (respectively 
$L^{2}(\R^{m})$ and $\ell^{2}((\Z/L)^{m}+k)$), it is easy to see the 
decomposition property which justifies (\ref{Bloch-dec}):
\begin{equation}
	\la f,g \ra_{L^{2}(\R^{m})} = \int_{[0,1/L)^{m}} \la f^{(k)}, 
	g^{(k)}\ra_{\Ltwok} dk
	\label{Bloch-dec2}
\end{equation}
An explicit formula for $f^{(k)}$, more direct than 
(\ref{def-fk}), is also computable with the aid of the Poisson 
summation formula:
\begin{equation}
	f^{(k)}(x) = \sum_{n\in\Z^{m}} \emi{Ln\cdot k} f(x+Ln)
\end{equation}
\smallskip
We can now proceed to the construction of a remarkable example of states 
satisfying the assumptions of the theorems.
Let a family of generalized \cs s for the Euclidean $2m$-dimensional 
phase space be given
\begin{equation}
	f_{(u,v)} := T(-u,v) f_{0}
	\label{cs-on-Rm}
\end{equation}
as constructed in \cite{p}, where $(u,v)\in\R^{2m}$ and $f_{0}\in 
L^{2}(\R^{m})$, usually a Gaussian centered at the origin.
According to our preparatory remark, we give the following
\begin{defn}
	{\rm\cite{dbg}} The family $f_{(u,v)}^{(k)}$ where $(u,v,k)\in X_{L}^{m}
	:= \TLm\times\R^{m} \times [0,1/L)^{m}$ constructed as above is called 
	a set of {\em \cs s on} $\LLm$.
	\label{cs-on-torus}
\end{defn}
We endow $X_{L}^{m}$ with the measure $d\theta (u,v,k) := du\,dv\,dk$ and 
check that they verify the hypothesis of the theorem.
\par
We shall work on the \Fou\ antitransform of $w_{\lambda}$, i.e. on the 
\Fou-Wigner \fn\ relative to the state $f_{\lambda}$.
\begin{eqnarray}
	&\pha& \int_{\LLm}dudv \int_{[0,1/L)^{m}}dk \left\la 
	f_{(u,v)}^{(k)}, \Tkex f_{(u,v)}^{(k)} \right\ra_{\Ltwok} = 
	\nonumber \\
	&=& \int_{\LLm}dudv \la f_{(u,v)},\Tex f_{(u,v)} 
	\ra_{L^{2}(\R^{m})} = \nonumber \\
	&=& \int_{\LLm}dudv \la T(-u,v)f_{0}, T(-u,v)\Tex f_{0} 
	\ra_{L^{2}(\R^{m})}\,\ei{(\eta\cdot v + \xi\cdot u)} =  \\
	&=& \int_{\LLm}dudv \la f_{0}, \Tex f_{0} \ra_{L^{2}(\R^{m})}\, 
	\ei{(\eta\cdot v + \xi\cdot u)} = \nonumber \\
	&=& \la f_{0}, \Tex f_{0} \ra_{L^{2}(\R^{m})} \,\delta_{\xi} 
	\delta(\eta) = \delta_{\xi}\delta(\eta), \nonumber 
\end{eqnarray}
which is another way to state (\ref{hyp}). The first step is justified 
by (\ref{Bloch-dec2}) and the third by the commutation relations in the 
Heisenberg group.
\par
Finally, this set of \cs s is perhaps the most important among the 
possible collections one could choose. As a matter of fact, such 
states are indeed introduced to be as {\em localized} as the Heisenberg 
principle permits, as clearly explained in \cite{dbg}. Had we 
performed a limit $\hbar\to 0$, exploiting the Wigner \fn\ as a measure 
of the degree of localization of a state, we would have seen that
\begin{equation}
	W_{f_{(u,v)}^{(k)},f_{(u,v)}^{(k)}}(p,q) \longrightarrow
	\delta(p-u)\delta(q-v) \mbox{ \ \ as } \hbar\to 0;
\end{equation}
where the dependence on $\hbar$ is implicit in the construction of 
$f_{(u,v)}^{(k)}$. This is why one can say that such a state is a good 
analogue of a point in the phase space. Therefore we see that the 
physical meaning of our quantum \erg\ properties gets clearer and more 
classical, of course, in the semiclassical regime.

\subsection{Proof of Lemma 4.1.} 
\label{appendix-pf-quant-evol}

Let us check that equality on all the matrix elements with 
respect to the standard basis of $\Ltwok$, $\{ e_{\alpha}^{(k)} 
\}_{\alpha\in\TLms}$ introduced in \S\ref{subs-quant}. It is 
easily computed that
\begin{equation}
	T^{(k)}(\eta,\xi) e_{\alpha}^{(k)} = \ei{\eta\cdot (\xi/2 + \alpha + 
	k)} e_{\alpha+\xi}^{(k)}.
	\label{Tealpha}
\end{equation}
If we now denote $c^{t}(p,q) := (c\circ\fLmt) = c(p,q+pt)$, we can 
compute its \Fou\ transform which turns out to be $\hat{c^{t}} 
(\eta,\xi) = \hat{c}(\eta - \xi t,\xi)$. Thus, substituting into 
(\ref{Opb1}) and changing variable,
\begin{equation}
	\Op(c^{t}) := \frac{1}{L^{m}} \sum_{\xi\in\TLms} \int_{\R^{m}} 
	\hat{c}(\eta,\xi)\, T(\eta + \xi t,\xi) \,d\eta.
	\label{Opbt}
\end{equation}
In order for the statement to hold for every $c$, it is a 
necessary and sufficient condition that $\forall\alpha,\gamma \in 
\TLms$
\begin{equation}
	\la e_{\alpha}^{(k)}, T(\eta + \xi t,\xi) e_{\gamma}^{(k)} \ra = 
	\la e_{\alpha}^{(k)}, \ei{t H_{m}} T(\eta,\xi) \emi{t H_{m}}
	e_{\gamma}^{(k)} \ra.
	\label{to-verify}
\end{equation}
Using (\ref{Tealpha}) we find on the r.h.s. 
\begin{equation}
	\la e_{\alpha}^{(k)}, T(\eta + \xi t,\xi) e_{\gamma}^{(k)} \ra =
	\ei{(\eta + \xi t) \cdot (\xi/2 + \gamma + k)} \, \delta_{\alpha, 
	\gamma+\xi}.
\end{equation}
Since $P^{(k)} e_{\alpha}^{(k)} = (\alpha + k) e_{\alpha}^{(k)}$, on 
the l.h.s. we have
\begin{eqnarray}
	&\pha& \la e_{\alpha}^{(k)}, \ei{t H_{m}} T(\eta,\xi) \emi{t H_{m}}
	e_{\gamma}^{(k)} \ra = \nonumber \\
	&=& e^{\pi i (\alpha + k)^{2} t} \, e^{-\pi i (\gamma + k)^{2} t} \,
	\ei{\eta\cdot (\xi/2 + \gamma + k)}\, \delta_{\alpha,\gamma+\xi} \\
	&=& \ei{((\xi/2 + \gamma + k) \xi t + (\xi/2 + \gamma + k)\eta)} \,
	\, \delta_{\alpha,\gamma+\xi}, \nonumber 
\end{eqnarray}
where we have substituted for $\alpha$ its value $\gamma+\xi$. This 
relation finally verifies (\ref{to-verify}).
\qed

\subsection{Convolutions over tori and over Euclidean spaces.}
\label{appendix-conv}

The purpose of this section is to clarify the meaning of the symbol 
$*_{L}$, indicating $q$-convolution over $\TLm$, when applied to \fn s 
that are in principle defined over larger sets, like the \fn s 
$b\circ\PLm$, for instance. Also, we want to understand how this is 
related to $*_{\infty}$, the ordinary convolution on $\R^{m}$, when 
$L \to\infty$.
\par
Since the arguments here are essentially descriptive, we specialize 
to one dimension, without loss of generality. If $f$ is a nice \fn\ 
defined on $\R$, denote by $f^{(L)}$, its {\em periodic restriction}, 
i.e. the \fn, {\em defined again on $\R$}, which is $L$-periodic and 
coincides with $f$ on $[-L/2,L/2)$. Then if $g$ is also a nice \fn\ 
on $\R$, we define
\begin{eqnarray}
	(f *_{L} g)(x) &:=& (f^{(L)} * g^{(L)})(x) = \int_{-L/2}^{L/2} 
	f^{(L)}(y) g^{(L)}(x-y)\, dy =  \nonumber \\
	&=& \int_{-L/2}^{L/2} f(y) g^{(L)}(x-y)\, dy. 
	\label{conv-1}
\end{eqnarray}
Thus, for instance, $(f *_{L} g)(x) \ne \int_{-L/2}^{L/2} f(y) 
g(x-y) dy$. As $L \to\infty$, however, we expect this to be 
approximately true, at least for a fixed $x\in\R$. As a matter of fact, 
requiring some properties of $f$ and $g$, one can prove a useful lemma. 
Call $S_{R} := [-R,R]$.
\begin{lemma}
	Suppose $f \in C_{0}^{\infty}(\R)$ and ${\rm supp}\, f \subseteq 
	S_{R}$. Assume also that $|g|$ vanishes monotonically at infinity. 
	Defining 
	\begin{displaymath}
		h(x) = (f *_{L} g - f *_{\infty} g)(x),
	\end{displaymath}
	then one has, for $L$ sufficiently large,
	\begin{displaymath}
		h(x) \: \left\{
		\begin{array}{ll}
		\le M (\, |g(-L/2)| + |g(L/2-R)| \,) & {\rm for}\ x\in 
		[-L/2,-L/2+R) \\
		= 0 & {\rm for}\ x\in [-L/2+R,L/2-R] \\
		\le M (\, |g(L/2)| + |g(-L/2+R)| \,) & {\rm for}\ x\in (L/2-R,L/2]
		\end{array}
		\right.
	\end{displaymath}
	where $M := R \max |f|$.
	\label{lemma-conv-app}
\end{lemma}
\par
{\sc Proof of Lemma \ref{lemma-conv-app}.} Take $L$ so large that 
$L/2 > R$ and $|g|$ is increasing in $(-\infty,-L/2+R]$ and decreasing in 
$[L/2-R,+\infty)$.
\par
Looking at (\ref{conv-1}) and recalling the hypothesis on $f$, we can write
\begin{equation}
	(f *_{L} g)(x) = \int_{S_{R}} f(y) g^{(L)}(x-y)\, dy.
\end{equation}
Hence
\begin{equation}
	h(x) = \int_{S_{R}} f(y) (g^{(L)} - g)(x-y)\, dy.
	\label{conv-2}
\end{equation}
Now, $g^{(L)}(x-y)$ coincides with $g^{(L)}(x-y)$ when $x-y \in 
S_{L/2}$, that is, when $y \in [x-L/2,x+L/2]$. So (\ref{conv-2}) is 
rewritten as
\begin{equation}
	h(x) = \int_{S_{R} \setminus (x+S_{L/2})} f(y) (g^{(L)} - g)(x-y)
	\, dy.
\end{equation}
It is easily seen that if $x \in [-L/2+R,L/2-R]$, then $S_{R} \subseteq 
(x+S_{L/2})$ such that $h(x) = 0$ and part of the claim is proved.
If $x \in [-L/2,-L/2+R)$, making the change of variable $z=x-y$, 
(\ref{conv-2}) gives
\begin{eqnarray}
	|h(x)| &=& \left| \int_{x-R}^{-L/2} f(x-z) (g^{(L)} - g)(z) dz \right| 
	\le \nonumber \\
	&\le& \max|f| \int_{-L/2-R}^{-L/2} (|g^{(L)}(z)| + |g(z)|) dz.
\end{eqnarray}
Since by definition, for $z< -L/2$, $g^{(L)}(z) = g(z+L)$, the 
monotonicity property of $g$ gives the first case in the statement of 
the lemma. The third case is of course analogous.
\qed

\subsection{Proof of Lemma 4.3.} 
\label{appendix-mix-tech}

Before even getting started, let us agree upon denoting,
throughout this proof, by $(p,q)$ all momentum-coordinate variables, 
be they defined on $\Lambda_{L}^{m}$ or on $\R^{2n}$ or on 
$\Lambda_{L}^{1}$. Notice that, in the proof of Theorem \ref{thm-mix}, 
we referred to $n$-dimensional variables as $(p',q')$.
\par 
The whole idea here is to realize that, if $n$ is fixed, the function 
$\Phi_{L}^{n}$ defined as in (\ref{phiLm-1}), gets closer and closer, 
in $\LLn$, to
\begin{eqnarray}
	\Phi_{\infty}^{n}(p,q) &=& e^{\beta p^{2}} \int_{\R^{n}} d\xi 
	\: e^{-\beta(p-\xi/2)^{2}} \, \ei{\xi\cdot q} =   \nonumber \\
	&=& e^{\beta p^{2}} \, e^{4\pi i p\cdot q}
	\left( \frac{4\pi}{\beta} \right)^{n/2} 
	\, e^{-(4\pi^{2}/\beta) q^{2}} =  \label{def-phi-inf}  \\
	&=& e^{\beta p^{2}} \, e^{4\pi i p\cdot q} \upsilon^{n}(q),
	\nonumber
\end{eqnarray} 
where $\upsilon(q) := \sqrt{4\pi/\beta}\: e^{-(4\pi^{2}/\beta) q^{2}}$. 
In the following we will use repeatedly the asymptotic estimate 
$\upsilon(L/2) = \OL$.
\par
Anyway, $\Phi_{\infty}^{n}$ is defined by the fact that it has the same 
\Fou\ spectrum of $\Phi_{L}^{n}$, suitably extended to all of $\R^{n}$.
If we denote by $\tilde{\cdot}$ the $q$-\Fou\ transform over $\TLn$, then
this amounts to say that, for $\xi\in (\TLn)^{*} = (\Z/L)^{n}$,
\begin{equation}
	\int_{\R^{n}} \Phi_{\infty}^{n}(p,q) \, \emi{\xi\cdot q} \,dq =
	e^{\beta p^{2}}\, e^{-\beta(p-\xi/2)^{2}} =: \tilde{\Phi}_{L}^{n}
	(p,\xi).
	\label{tilde-phi-inf}
\end{equation}
So the best candidate 
for $\gamma_{\infty}$ is $\beta *_{\infty} \Phi_{\infty}^{n}$. The 
symbol $*_{\infty}$ designates the $q$-convolution over 
$\R^{n}$, as explained in \S\ref{appendix-conv}. First of all, 
such a $\gamma_{\infty}$ verifies (\ref{lemma-t-two}): 
this is a consequence of the fact that $\int_{\R^{n}} 
\Phi_{\infty}^{n} = 1$, which is easily verified. Let us proceed to the 
proof of (\ref{lemma-t-one}). 
\smallskip\par
Recalling the warning in \S\ref{appendix-conv}, the main 
inequality will be
\begin{eqnarray}
	&\pha& \| \gamma_{L} - \gamma_{\infty} \|_{L^{2} (\LLn,\mu_{L}^{n})} 
	= \| \beta *_{L} \Phi_{L}^{n} - \beta *_{\infty} \Phi_{\infty}^{n} 
	\|_{L^{2} (\LLn,\mu_{L}^{n})} \le \label{tech-fund} \\
	&\le& \| \beta *_{L} (\Phi_{L}^{n} - \Phi_{\infty}^{n}) \|_{L^{2} 
	(\LLn,\mu_{L}^{n})} + \| \beta *_{L} \Phi_{\infty}^{n} - 
	\beta *_{\infty} \Phi_{\infty}^{n} \|_{L^{2}(\LLn,\mu_{L}^{n})}.
	\nonumber
\end{eqnarray}
The leftmost term can be worked out pointwise, using
(\ref{def-phi-inf}) and Lemma \ref{lemma-conv-app} of  
\S\ref{appendix-conv}. If $(p,q) \in \LLn$,
\begin{equation}
	|\, ( \beta *_{L} \Phi_{\infty}^{n} - \beta *_{\infty} 
	\Phi_{\infty}^{n}) (p,q) \,| \le M \upsilon^{n} 
	\left( \frac{L}{2} -R \right),
	\label{tech-00}
\end{equation}
where $M \simeq \max (|\beta(p,q)| e^{\beta p^{2}})$ and $R$ is the 
radius of the ball containing supp$\,\beta$. Now, since 
$\mu_{L}^{n}$ is a probability measure,
\begin{equation}
	\| \beta *_{L} \Phi_{\infty} - \beta *_{\infty} \Phi_{\infty} 
	\|_{L^{2}(\LLn,\mu_{L}^{n})} = \OL.
	\label{tech-01}
\end{equation}
To work out the other term in (\ref{tech-fund}) we employ the 
ideas stated at the beginning of this section about $\Phi_{\infty}^{n}$. 
We start by applying the well-known convolution inequality to our case. 
We have
\begin{equation}
	\int_{\TLn} | \beta *_{L} (\Phi_{L}^{n} - \Phi_{\infty}^{n}) 
	(p,q)|^{2} \, dq \le \| \beta(p,\cdot) \|_{L^{1}(\TLn,dq)}^{2} 
	\: \| (\Phi_{L}^{n} - \Phi_{\infty}^{n}) (p,\cdot) 
	\|_{L^{2}(\TLn,dq)}^{2}. 
	\label{tech-1}
\end{equation}
Notice that $\| \beta(p,\cdot) \|_{L^{1}(\TLn)} = \| \beta(p,\cdot) 
\|_{L^{1}(\R^{n})}$, since $\beta$ is compactly supported. We also 
see that
\begin{equation}
	\| (\Phi_{L}^{n} - \Phi_{\infty}^{n}) (p_{1}, \ldots, p_{n},\cdot) 
	\|_{L^{2}(\TLn,dq)}^{2} = \prod_{i=1}^{n} \| (\Phi_{L} -
	\Phi_{\infty}) (p_{i},\cdot) \|_{L^{2}(T_{L},dq_{i})}^{2},
	\label{tech-2}
\end{equation}
since $\Phi_{L}^{n}$ and $\Phi_{\infty}^{n}$ are completely 
factorizable: we call, obviously, $\Phi_{L}$ and $\Phi_{\infty}$ their 
one-dimensional versions, on which we are immediately going to work. 
As anticipated at the beginning of this section, we will be a little 
imprecise and use again the label $(p,q)$ for $(p_{i},q_{i})$.
Now
\begin{equation}
	\| (\Phi_{L} - \Phi_{\infty}) (p,\cdot) \|_{L^{2}(T_{L})}^{2} = 
	\frac{1}{L} \sum_{\xi\in(\Z/L)} | \tilde{\Phi}_{L} (p,\xi) - 
	\tilde{\Phi}_{\infty} (p,\xi) |^{2},
	\label{tech-3}
\end{equation}
with $\tilde{\Phi}_{\infty} (p,\xi) = \int_{-L/2}^{L/2} 
\Phi_{\infty} (p,q) \emi{\xi q} dq$. Looking back at 
(\ref{tilde-phi-inf}) and using definition (\ref{def-phi-inf}), 
we can write
\begin{eqnarray}
	& \pha & | \tilde{\Phi}_{L} (p,\xi) - \tilde{\Phi}_{\infty} (p,\xi) |
	= \nonumber  \\ 
	& = & \left| \int_{\R \setminus T_{L}} \Phi_{\infty}(p,q) \, \emi{\xi q} 
	\,dq \right| =  \nonumber \\
	& = & 2 e^{\beta p^{2}} \left| \int_{L/2}^{+\infty} \upsilon(q) \, 
	\cos(2\pi(\xi -2p) q)\, dq \right| =  \label{tech-4} \\
	& = & 2 e^{\beta p^{2}} \left| \left[ \frac{\upsilon(q) \, \sin(2\pi(\xi 
	-2p) q)} {2\pi (\xi -2p)} \right]_{L/2}^{+\infty} - 
	\int_{L/2}^{+\infty} \frac{\upsilon'(q) \, \sin(2\pi(\xi -2p) q)} 
	{2\pi (\xi -2p)}\, dq \right| \le  \nonumber \\
	& \le & e^{\beta p^{2}} g(\xi -2p) \upsilon(L/2), \nonumber
\end{eqnarray}
where $g(x)$ is a continuous \fn\ defined on $\R$ behaving like 
$|x|^{-1}$ for large values of $x$. Note that, as $L\to\infty$, 
$(1/L) \sum_{\xi \in (\Z/L)} g^{2}(\xi -2p) \to \int_{\R} 
g^{2}(\xi) d\xi =: K$, {\em uniformly} for $p$ in a compact set.
So, looking at (\ref{tech-1}), and merging up (\ref{tech-2}), 
(\ref{tech-3}) and (\ref{tech-4}), we have
\begin{eqnarray}
	&\pha& \| \beta *_{L} (\Phi_{L}^{n} - \Phi_{\infty}^{n}) \|_{L^{2} 
	(\LLn,\mu_{L}^{n})}^{2} = \nonumber  \\
	&=& \frac{1}{L^{n}} \int_{\R^{n}} \| \beta *_{L} (\Phi_{L}^{n} - 
	\Phi_{\infty}^{n} (p,\cdot) \|_{L^{2}(\TLn,dq)}^{2} \: 
	d\breve{\mu}^{n} (p) \le \label{tech-5}  \\
	&\le& \left( \frac{(K+1) \upsilon^{2}(L/2)}{L^{n}} \right)^{n} 
	\int_{\R^{n}} \| \beta(p,\cdot) \|_{L^{1}(\R^{n})}^{2} e^{2\beta p^{2}} 
	\: d\breve{\mu}^{n} (p) = \OL, \nonumber
\end{eqnarray}
since $\beta$ has compact support. Inserting (\ref{tech-01}) and 
(\ref{tech-5}) into the fundamental inequality
(\ref{tech-fund}) the proof is completed.
\qed

\end{document}